\documentclass[useAMS,usenatbib]{mn2e}
\usepackage{amssymb}
\usepackage{amsmath}
\usepackage{graphicx}
\usepackage{BibDef}
\usepackage[bookmarks=false]{hyperref}
\usepackage{tabularx}
\usepackage{ulem}
\def\lsim{\mathrel{\rlap{\lower 3pt \hbox{$\sim$}} \raise 2.0pt \hbox{$<$}}}
\def\gsim{\mathrel{\rlap{\lower 3pt \hbox{$\sim$}} \raise 2.0pt \hbox{$>$}}}
\def\msun{\rm {M_{\large \odot}}}

\title[Post-Newtonian MBH Dynamics] {Post-Newtonian  Evolution of Massive Black Hole Triplets in Galactic Nuclei: 
I.Numerical Implementation and Tests}
\author[Bonetti et al.]{Matteo Bonetti$^{1,2}$, Francesco Haardt$^{1,2}$, Alberto Sesana$^3$ \& Enrico Barausse$^{4,5}$\\
$^1$DiSAT, Universit\`a degli Studi dell'Insubria, Via Valleggio 11, 22100 Como, Italy\\
$^2$INFN, Sezione di Milano-Bicocca, Piazza della Scienza 3, 20126 Milano, Italy\\
$^3$School of Physics and Astronomy, University of
Birmingham, Edgbaston, Birmingham B15 2TT, United Kingdom\\
$^4$CNRS, UMR 7095, Institut d'Astrophysique de Paris, 98 bis Bd Arago, 75014 Paris, France\\
$^5$Sorbonne Universit\'es, UPMC Univesit\'e Paris 6, UMR 7095, Institut d'Astrophysique de Paris, 98 bis Bd Arago, 75014 Paris, France
}

\begin{document}

\date{~}

\pagerange{\pageref{firstpage}--\pageref{lastpage}} \pubyear{2015}

\maketitle

\label{firstpage}

\begin{abstract} 
Massive black-hole binaries (MBHBs) are thought to be the main source of gravitational waves (GWs) in the low-frequency domain surveyed by ongoing and forthcoming Pulsar Timing Array campaigns and future space-borne missions, such as {\it eLISA}. However, many low-redshift MBHBs in realistic astrophysical environments may not reach separations small enough to allow significant GW emission, but rather stall on (sub)pc-scale orbits. This ``last-parsec problem" can be eased by the appearance of 
a third massive black hole (MBH) -- the ``intruder" -- whose action can force, under certain conditions, the inner MBHB on a very eccentric orbit, hence allowing intense GW emission eventually leading to coalescence. 
A detailed assessment of the process, ultimately driven by the induced Kozai-Lidov oscillations of the MBHB orbit, requires a general relativistic treatment and the inclusion of external factors, such as  the Newtonian precession of the intruder orbit in the galactic potential and its hardening by scattering off background stars. 
In order to tackle this problem, we developed a three-body Post-Newtonian (PN) code framed in a realistic galactic potential, 
including both non-dissipative 1PN and 2PN terms, and dissipative terms such as 2.5PN effects, orbital hardening of the outer binary, and the effect of the dynamical friction on the early stages of the intruder dynamics. 
In this first paper of a series devoted at studing the dynamics of MBH triplets from a cosmological perspective, we describe, test and validate our code.  
\end{abstract}

\section{Introduction}

Massive black holes (MBHs), ubiquitous in the nuclei of nearby galaxies \citep[see][and references therein]{Kormendy2013}, are recognised to be a fundamental ingredient in the process of galaxy formation and evolution along the cosmic history of the Universe. In the bottom-up hierarchical clustering of dark matter overdensities predicted by $\Lambda$CDM cosmology, the notion that MBHs were common in galaxy nuclei at all epochs leads to the inevitable conclusion that a large number of massive black hole binaries (MBHBs) did form during the build-up of the large scale structure (Begelman et al. 1980). 

MBHBs are expected to be the loudest sources of gravitational radiation in the nHz-mHz frequency range \citep{Haehnelt1994,Jaffe2003,Wyithe2003,Enoki2004,Sesana2004,Sesana2005,Jenet2005,Rhook2005,2012MNRAS.423.2533B,2016PhRvD..93b4003K}, a regime partially covered by the planned {\it eLISA} space-born interferometer \citep{2013arXiv1305.5720C}, and by existing 
Pulsar Timing Array (PTA) experiments \citep{2016MNRAS.458.3341D,2015ApJ...813...65T,2016MNRAS.455.1751R,2016MNRAS.458.1267V}.

One of the key open questions that determine the observability of MBHBs by {\it eLISA} and PTAs is whether these systems merge at all within a Hubble time, an issue often referred to as  ``the last-parsec problem''. Whenever two galaxies merge, it is understood that the MBHs they host will fall toward the center of the newly formed galaxy as a result of dynamical friction against stars and gas. When the relative velocity of the two MBHs exceeds that of the background stars, 
dynamical friction becomes ineffective at further shrinking their separation, and the two MBHs form a bound binary. 

At this stage, single three-body interactions of the MBHB with background stars transfers energy from the MBHB (which will shrink as a result) to the stars \citep[which may be ejected from the MBHB surroundings or even from the galaxy; see, e.g.,][]{Sesana2008}. After a first phase of fast orbital shrinking, the binary hardens at 
constant rate once it reaches a separation of the order of the so-called hardening radius, $a_h\sim Gm_2/4\sigma^2$ \citep[here $m_2$ is the mass of the secondary, and $\sigma$ is the stellar velocity dispersion, see][]{Quinlan1996}. 
Typically $a_h\sim 1$ pc for MBHs with masses $\sim 10^8 \msun$, a value significantly larger than the separation $a_{\rm gr}\sim 10^{-2}$ pc at which gravitational wave (GW) emission alone can drive the binary to coalescence within a Hubble time. 

Stellar hardening, however, is only efficient as long as stars with orbits intersecting the MBHB (the ``loss-cone" in the energy-angular momentum parameter space of stars) do actually exist. Since in the process stars are ejected by the slingshot mechanism, new stars need  to continuously replenish the loss-cone~\citep{yu}. Loss-cone replenishment may simply happen due to diffusion of stars in energy-angular momentum space, which occurs on the stellar relaxation time. Since the latter is typically longer than the Hubble time for galaxies hosting MBHs larger than $\sim 10^9 M_\odot$, additional mechanisms are needed to replenish the loss-cone if MBHBs are to coalesce in the most massive systems. Proposed ways to enhance stellar diffusion are for instance a possible triaxiality in the galactic potential  -- following, e.g., from recent mergers~\citep{yu,lastPc1,2014CQGra..31x4002V,lastPc2,2015ApJ...810...49V} -- or galactic rotation~\citep{2015arXiv150506203H}. This may lead MBHBs to merge within a timescale of a few Gyr or less, depending on the binary eccentricity \citep{Sesana2015}.

If a significant amount of gas is present in the galactic nucleus, the evolution of the binary down to GW dominated separations may be faster, because MBHBs are expected to undergo planetary-like migration, which may drive the MBHB to coalescence within $\sim 10^7$--$10^8$ yr~\citep{bence,Colpi:2014poa}. However, this is not expected to be a major effect for most low-redshift galaxies hosting MBHs with masses in the PTA range ($\sim 10^9 M_\odot$), since those galaxies are typically gas-poor. Gas interactions, however, may be important for {\it eLISA}, i.e., for MBHBs in the mass range $10^4$--$10^7 M_\odot$ \citep{2007MNRAS.379..956D}.

Finally, a third major process that could solve the last parsec problem is provided by triple MBH interactions, which might occur when a MBHB stalled at separations $\lesssim a_{\rm h}$ (because of the lack of sufficient gas and inefficient loss-cone replenishment) interacts with a third MBH -- the ``intruder'' -- carried by a new galaxy merger.
More specifically, these hierarchical triplets  -- i.e., triple systems where the hierarchy of orbital separations allows one to 
define an inner and an outer binary, the latter consisting of the intruder and the center of mass of the former -- may undergo Kozai-Lidov (K-L) oscillations \citep{Kozai1962,Lidov1962}. These particular resonances take place if the intruder is on a highly inclined orbit with respect to the inner binary, and tend to secularly increase the eccentricity of the inner binary, eventually driving it to coalescence.

The standard method for the investigation of the K-L mechanism leverages on a perturbative approach of the secular restricted three-body problem, in which the Hamiltonian is expanded as power series in terms of the inner to outer semi-major axis ratio. In the original works of Kozai and Lidov only the first term in the expansion, i.e., the quadrupole term proportional to $(a_{\rm in}/a_{\rm out})^2$, was taken into account. Higher order terms result in new phenomenology \citep{Ford2000}, which has been recently reviewed in \cite{Naoz2016}. 
Several other works have been focusing on the study of the K-L mechanism in different astrophysical contexts including: planetary dynamics \citep{Holman1997,Katz2011,Naoz2012,Naoz2013}, interactions of stellar size objects in globular clusters \citep{Antonini2016, Antognini2016} and around MBHs, and triple MBH systems \citep{Blaes2002,Iwasawa2006,Iwasawa2008,Hoffman2007,Pau2010}. 
 
A first study devoted to MBH triplets was presented in \cite{Blaes2002}, where the secular (i.e., orbit-averaged) Hamiltonian of a hierarchical triplet was expanded up to octupole order. The authors also included secular terms to account for relativistic precession and GW losses, but neglected the second post Newtonian (2PN) term. Through numerical integration of the orbit averaged equations of motion, \citet{Blaes2002} proved that the merger time-scale of a MBHB can be reduced by up to an order of magnitude.
A Newtonian direct N-body simulation including a MBH triplet {\it and} background stars was later presented by \citet{Iwasawa2006,Iwasawa2008}, who included GW losses for the three MBHs. The authors found that K-L oscillations and/or strong resonant interactions can greatly increase the eccentricity of the innermost binary, leading it to coalesce.

\citet{Hoffman2007} embarked in a systematic study of MBH triplets embedded in a stellar background. Triplets were initialised 
from astrophysically and cosmologically motivated initial conditions, and the dynamics was purely Newtonian but with the introduction of a recipe to account for gravitational radiation. The authors found that three-body interactions can enhance the rate of MBHB coalescences, suggesting the production of burst-like GW signals due to the high eccentricities of the systems. 

More recently, \citet{Pau2010} framed the dynamics of triplets in the specific context of GW detection with PTAs. Using a large library of systems from \citet{Hoffman2007}, they studied the phenomenology of bursts observable by PTAs, by assuming an arbitrary fraction of MBHBs driven by triple interactions.

In the present paper we focus on the dynamics of MBH triplets, similarly in spirit to \citet{Hoffman2007}. A major improvement is the introduction of all relativistic corrections up to 2.5PN order, which are crucial when dealing with K-L oscillations, since pericenter precession can strongly inhibit the mechanism \citep{Holman1997}. We will also extend our investigation to systems spanning a much larger parameter space in terms of MBH masses and mass ratios. This is our first step of a planned series of four papers devoted to a detailed analysis of the post-Newtonian evolution of MBH triplets in galactic nuclei. Here we present and test our numerical integrator. In the second paper of the series we will discuss the astrophysical conditions for the formation of MBH triplets, while in the third step we will frame the whole picture in a cosmological context, adopting a semi-analytical model of galaxy and black hole co-evolution. In a final fourth paper we will compute waveforms and make predictions for current and future low frequency GW searches, such as PTAs and {\it eLISA}.

\section{Three bodies in a stellar background}

Three-body Newtonian dynamics is largely textbook matter. Here we first briefly summarise the post-Newtonian approach to the general relativistic problem, and then we present in detail our original method to include the effects of the stellar environment on a  
hierarchical MBH triplet.

\subsection{PN equations of motion}

The Hamiltonian for a triple system of non-spinning bodies is given, in schematic form and through 2.5PN order\footnote{The PN
approximation expands the dynamics perturbatively 
in the ratio $v/c$, $v$ being the binary's relative velocity~\citep{WillPN1993}. A term suppressed by a factor $(v/c)^{2 n}$
with respect to the leading (Newtonian) order is said to be of $n$PN order.}, by 
\begin{equation}\label{eq:hamiltonian}
{H} = {H}_0 + \dfrac{1}{c^2}{H}_1 + \dfrac{1}{c^4}{H}_2 + \dfrac{1}{c^5}{H}_{2.5} + O\left(\dfrac{1}{c^6}\right)\,.
\end{equation}
%
Each PN order, through its corresponding Hamiltonian, introduces different relativistic corrections to the standard Newtonian laws of motion, and therefore a qualitatively different dynamics. Even powers of $c^{-1}$ represent conservative terms, while odd powers are dissipative terms. In particular, $H_0$, $H_1$ and $H_2$ are functions
of the positions and conjugate momenta of the three bodies, i.e., 
$\vec{x}_{\alpha}$ and $ \vec{p}_{\alpha}$ (with $\alpha=1,2,3$), while
the dissipative 2.5PN Hamiltonian ${H}_{2.5}$ must also depend explicitly on time
to account for the leading-order backreaction of GW emission onto the triplet's dynamics (this is because
otherwise $dH/dt=\partial H/\partial t=0$, which would imply energy conservation).

The equations of motion for the $\alpha$-th body then take the usual form
\begin{align}\label{eq:EOM1}
\dot{\vec{x}}_{\alpha} &= \sum_n \dfrac{1}{c^{2 n}} \dfrac{\partial H_n}{\partial \vec{p}_{\alpha}}\\ 
\dot{\vec{p}}_{\alpha} &= -\sum_n  \dfrac{1}{c^{2 n}}\dfrac{\partial H_n}{\partial \vec{x}_{\alpha}}\,.
\label{eq:EOM2}
\end{align}
Explicit expressions for the Hamiltonian are given by  \citet{1987PhLA..123..336S}, \citet{Lousto2008} and \citet{Galaviz2011} \citep[see also][for higher-order terms]{Konigsdorffer2003}, and are reported in the appendix at Newtonian, 1PN and 2PN order. The dissipative 2.5PN Hamiltonian is instead slightly trickier
and we give it explicitly here:
\begin{equation}\label{eq:ham2.5}
H_{2.5} = \dfrac{G}{45}\chi^{ij}(\vec{x}_{\alpha},\vec{p}_{\alpha})\,\dot{\chi}_{ij}(\vec{x}_{\alpha'},\vec{p}_{\alpha'}),
\end{equation}
where
\begin{align}\label{eq:chi}
\chi_{ij}(\vec{x}_{\alpha},\vec{p}_{a}) = &\sum_{\alpha} \dfrac{2}{m_{\alpha}}\left(|\vec{p}_{\alpha}|^2\delta_{ij}-3p_{\alpha,i}p_{\alpha,j}\right)\notag
\\&+\sum_{\alpha}\sum_{\beta\neq \alpha} \dfrac{G m_{\alpha} m_{\beta}}{r_{\alpha\beta}}\left(3n_{\alpha\beta,i}n_{\alpha\beta,j}-\delta_{ij}\right)
\end{align}
and
\begin{align}\label{eq:chi_dot}
\dot{\chi}_{ij}(\vec{x}_{\alpha'},&\vec{p}_{\alpha'}) =\notag\\
& \sum_{\alpha'}\dfrac{2}{m_{\alpha'}}\Bigl[2(\dot{\vec{p}}_{\alpha'}\cdot\vec{p}_{\alpha'})\delta_{ij} - 3(\dot{p}_{\alpha'i}p_{\alpha'j} + p_{\alpha'i}\dot{p}_{\alpha'j})\Bigr]\notag\\
&+\sum_{\alpha'}\sum_{\beta'\neq \alpha'}\dfrac{G m_{\alpha'}m_{\beta'}}{r^2_{\alpha'\beta'}} \Bigl[ 3(\dot{r}_{\alpha'\beta'i}n_{\alpha'\beta'j} + n_{\alpha'\beta'i}\dot{r}_{\alpha'\beta'j})\notag\\
& + (\vec{n}_{\alpha'\beta'}\cdot \dot{\vec{r}}_{\alpha'\beta'})(\delta_{ij}-9 n_{\alpha'\beta'i}n_{\alpha'\beta'j})   \Bigr].
\end{align}
Here, Latin indices label 3-dimensional vector components (e.g., 
$x_{\alpha i}$ indicates the $i$-th position coordinate of the $\alpha$-th body), $\delta_{ij}$ is the usual 
Kroneker delta, and we have defined 
\begin{align}\label{eq:varaibles3}
\vec{r}_{\alpha\beta} &= \vec{x}_{\alpha} - \vec{x}_{\beta}\notag\\
r_{\alpha\beta} &= |\vec{r}_{\alpha\beta}|\notag\\
\vec{n}_{\alpha\beta} &= \frac{\vec{r}_{\alpha\beta} }{ r_{\alpha\beta}}.
\end{align}
Primed quantities denote retarded variables that are not subject to the derivative operators in eqs.~\ref{eq:EOM1} and~\ref{eq:EOM2}. Primed and unprimed variables are then identified once the derivatives in the equations of motion (eqs.~\ref{eq:EOM1} and~\ref{eq:EOM2}) have been calculated. This implicitly makes $H_{2.5}$ time-dependent, as expected. In more detail, the 2.5PN contribution to the equations of motion reads
\begin{align}\label{eq:EOM2.5}
(\dot{\vec{x}}_{\alpha})_{2.5} &=\dfrac{1}{c^5}\dfrac{\partial H_{2.5}}{\partial \vec{p}_{\alpha}}\notag\\&= \dfrac{G}{45 c^5} \dot{\chi}_{ij}(\vec{x}_{\alpha},\vec{p}_{\alpha}; \dot{\vec{x}}_{\alpha},\dot{\vec{p}}_{\alpha})\dfrac{\partial}{\partial\vec{p}_{\alpha}}\chi^{ij}(\vec{x}_{\alpha},\vec{p}_{\alpha})\\
(\dot{\vec{p}}_{\alpha})_{2.5} &=-\dfrac{1}{c^5}\dfrac{\partial H_{2.5}}{\partial \vec{x}_{\alpha}}\notag\\&= -\dfrac{G}{45 c^5} \dot{\chi}_{ij}(\vec{x}_{\alpha},\vec{p}_{\alpha}; \dot{\vec{x}}_{\alpha},\dot{\vec{p}}_{\alpha})\dfrac{\partial}{\partial\vec{x}_{\alpha}}\chi^{ij}(\vec{x}_{\alpha},\vec{p}_{\alpha}).
\end{align} 
Note that in order to have terms not higher than 2.5PN, the time derivatives of the positions and conjugate momenta that appear in the function $\dot{\chi}_{ij}$ should be replaced by their  Newtonian limits, i.e., $\dot{\vec{x}}_{\alpha}\to\partial H_0/\partial {\vec{p}}_{\alpha}$ and $\dot{\vec{p}}_{\alpha}\to-\partial H_0/\partial {\vec{x}}_{\alpha}$. 

\subsection{Hardening in a fixed stellar background}

One of the key ingredients in the dynamical evolution of a hierarchical MBH triplet in a realistic post-merger situation is the hardening of the outer binary. Ambient stars, unbound to the outer binary, are expelled by the gravitational slingshot, carrying away the MBHB energy and angular momentum. As a result, the MBHB orbit gets tighter and more eccentric. 

As described in \citet{Quinlan1996}, the binary evolution in an isotropic fixed background of stars of density $\rho$ and one-dimensional velocity dispersion $\sigma$ can be expressed in terms of the dimensionless hardening rate $H$ and eccentricity growth rate $K$ as
\begin{align}
\dot{a} &= -a^2\dfrac{G\rho H}{\sigma},
\label{eq:a_evolution}\\
\dot{e} &= a\dfrac{G\rho H K}{\sigma}.
\label{eq:e_evolution}
\end{align}
\cite{Sesana2015} demonstrated that these equations are appropriate to describe the hardening in a galaxy merger remnant, provided that $\rho$ and $\sigma$ are defined at the binary influence radius, i.e., the radius containing twice the binary mass in stars. $H$ and $K$ can be derived from detailed three-body scattering experiments as in \citet{Sesana2006}, who provided numerical fits to $H$ and $K$ for various combinations of MBH mass ratios, eccentricities and separations. 

Our aim here is to include the hardening of the outer binary in the three-body dynamics described in the previous subsection. The methodology we employ relies on a velocity-dependent (hence dissipative) fictitious force, which is tuned to provide an orbital averaged decay and an eccentricity growth consistent with eqs.~\ref{eq:a_evolution} and~\ref{eq:e_evolution}. It is worth stressing that it is exactly the hierarchical nature of the triplets we consider that allows us to treat the hardening of the outer binary following \citet{Quinlan1996} and \citet{Sesana2006}. The hardening binary is, in our case, formed by the intruder and by the center of mass of the inner binary. 

Under the assumption of a small dissipative force, the rates of change of the orbital elements can be derived by standard perturbation theory in the framework of celestial mechanics. We start off by considering the perturbed two-body problem, i.e., 
\begin{equation}
\dfrac{d^2\vec{r}}{dt^2} = -\dfrac{G M}{r^3}\vec{r} + \vec{\delta}, 
\label{eq:2body_pert}
\end{equation}
where $M$ is the binary total mass, $\vec{r}$ the relative separation, and in the most general case the extra-acceleration $\vec{\delta}$ is a generic function of position, velocity and time. 
Given the planar geometry of the Keplerian problem, $\vec{\delta}$ can be projected along three directions, i.e., on the orbital plane along the radial direction ($S$) and normal to it ($T$), and along the direction orthogonal to the orbital plane ($W$):
\begin{equation}
S = \dfrac{\vec{\delta}\cdot\vec{r}}{r}, \quad T = \dfrac{\vec{\delta}\cdot(\vec{h}\times\vec{r})}{h r}, \quad W = \dfrac{\vec{\delta}\cdot\vec{h}}{h},
\label{eq:STW}
\end{equation}
where $\vec{h}$ is the binary's angular momentum per unit mass. 

During the hardening phase, the orientation of the orbital plane of a MBHB undergoes a random walk \citep{Merritt2002}. 
Though in an triaxial or axisymmetric 
potential (e.g., when the global rotation of the stellar bulge is important) the effect is relevant \citep[see][]{Gualandris2012}, 
in the case of a spherically symmetric stellar distribution as the one we consider here (see next \S2.3), the random orientation of the orbital plane is $\lsim 10^\circ$.  
We therefore neglect the re-orientation effect during the hardening phase, i.e., we set $W=0$.

While the energy per unit mass in the unperturbed Keplerian problem  is a constant of motion, the dissipative force is responsible for its variation in time, i.e., 
$\dot{E}= \vec{\delta}\cdot\vec{v}=S v_r + T v_t$, 
where $v_r$ and $v_t$ are the radial and tangential velocity, respectively.  In terms of 
eccentricity $e$ and orbital true anomaly $\nu$, the velocity components are written as
\begin{align}
v_r &= \dfrac{G M}{h} e \sin\nu ,\notag\\
v_t &= \dfrac{G M}{h} (1 + e \cos\nu ),
\end{align}
so that the energy variation becomes
\begin{equation}
\dot{E}= \left[ S e \sin\nu + T (1 + e \cos\nu ) \right] \dfrac{G M}{h}.
\label{eq:power_comp_f}
\end{equation}
Finally, from $h = \sqrt{G M a (1-e^2)}$, eq.~\ref{eq:power_comp_f} gives the variation rate of the semi-major axis $a$:
\begin{equation}
\dot{a}= \dfrac{2 a^2}{G M}\dot{E} = 2\sqrt{\dfrac{a^3}{G M(1-e^2)}}\left[ S e \sin\nu + T (1 + e \cos\nu ) \right].
\label{eq:a_variation}
\end{equation}

\begin{figure}
   \includegraphics[width=0.47\textwidth]{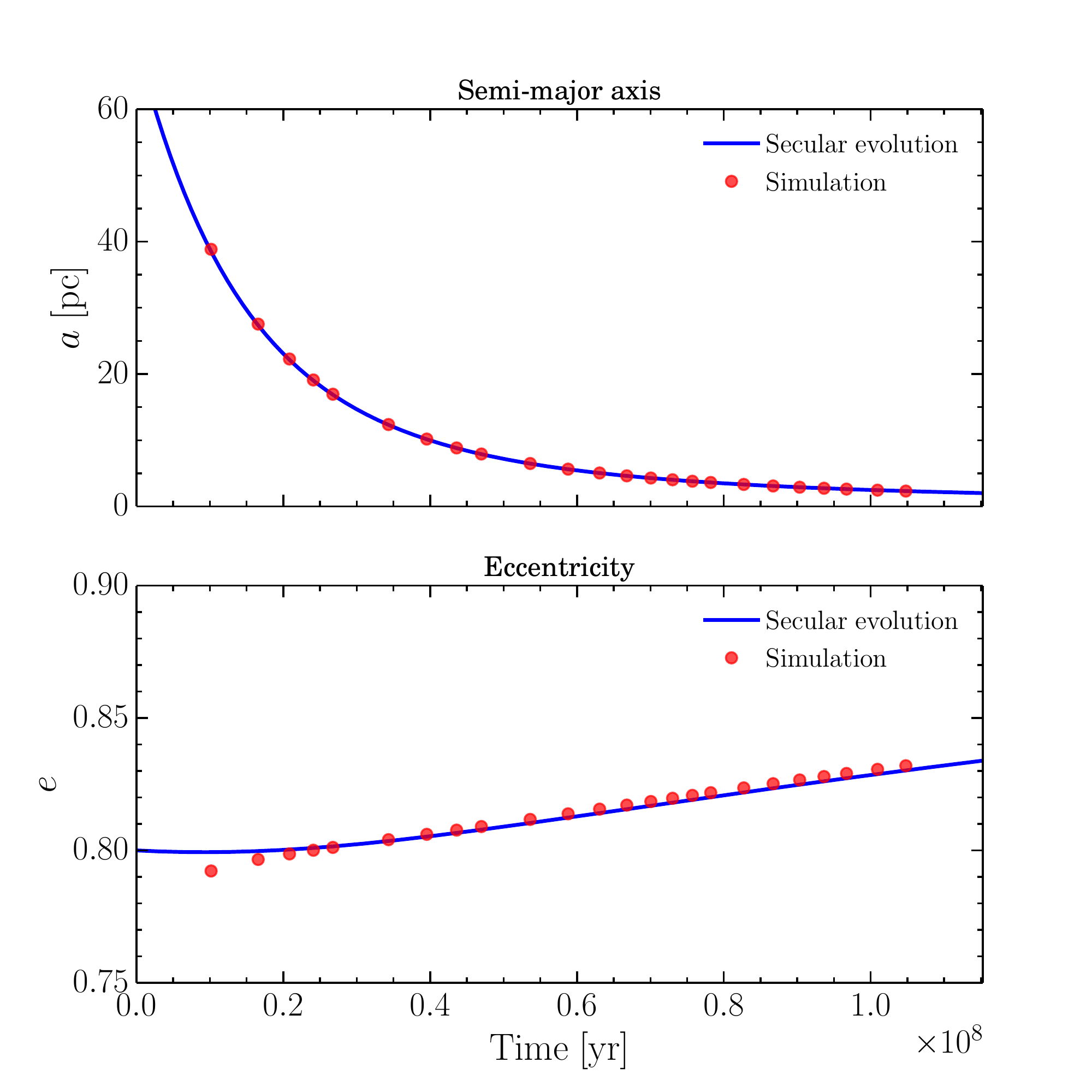}
   \caption{Semi-major axis ({\it upper panel}) and eccentricity ({\it lower panel}) evolution of a MBH binary scattering off background stars. Our implementation of the hardening process (shown as filled dots) is compared to the results of the scattering experiments of \citet{Sesana2006}, shown as solid lines.}
\label{fig:hardening}
\end{figure}

The perturbing force changes also the angular momentum according to the torque exerted. In our specific case (i.e., $W=0$), the angular momentum varies only in magnitude according to
\begin{equation}
\dot h = \dfrac{G M}{2 h}\left[ (1-e^2)\dot{a} - 2ae\dot{e} \right]=rT.
\label{eq:torque}
\end{equation}

By substituting $\dot{a}$ from eq.~\ref{eq:a_variation} and expressing $r$ in terms of $a$, $e$ and $\nu$, i.e.,  
\begin{equation}
r=\frac{a(1-e^2)}{1+e\cos\nu}, 
\end{equation}
we finally obtain the variation rate of the eccentricity,
\begin{equation}
\dot{e} = \sqrt{\dfrac{(1-e^2)a}{G M}}\left[S \sin\nu + T\dfrac{2\cos\nu + e(1+\cos^2\nu)}{1+e\cos\nu} \right].
\label{eq:e_variation}
\end{equation}

Eqs.~\ref{eq:a_variation} and~\ref{eq:e_variation} represent instantaneous variation rates, that need to be compared to the orbit averaged rates derived from scattering-experiment results (eqs.~\ref{eq:a_evolution} and~\ref{eq:e_evolution}). We then need to perform an orbit 
average of eqs.~\ref{eq:a_variation} and~\ref{eq:e_variation}, i.e.,  
\begin{align}\label{eq:a_av}
<\dot{a}> &= \dfrac{1}{P_{\rm orb}} \int_{0}^{P_{\rm orb}}dt\,  \dot{a},\\
<\dot{e}> &= \dfrac{1}{P_{\rm orb}} \int_{0}^{P_{\rm orb}}dt\, \dot{e}.\label{eq:e_av}
\end{align}
Note that, in order to numerically compute the above integrals, the integration over time must be substituted with an integration over the true anomaly. Starting from Kelper's equation, straightforward but rather long calculations give  
\begin{equation}\label{eq:dt}
dt = \sqrt{\dfrac{a^3}{G M}}\,\dfrac{(1-e^2)^{3/2}}{1+e \ \cos^2\nu}d\nu, 
\end{equation}
where we assumed negligible variations of both $a$ and $e$ along a single orbit.

Next, we need an appropriate form for $\vec{\delta}$ (i.e., appropriate $S$ and $T$), such that its orbit-averaged action produces orbital variations matching eqs.~\ref{eq:a_evolution} and~\ref{eq:e_evolution}. We assume that $\vec{\delta}$ is the sum of two terms, one orthogonal to $\vec{v}$ and one parallel to $\vec{v}$: 
\begin{equation}\label{eq:acc_vec}
\vec{\delta} = A \dfrac{\vec{v}\cdot\vec{r}}{r} \Biggl[ \dfrac{\vec{r} - (\vec{r}\cdot\vec{v})\vec{v}/v^2}{\sqrt{r^2-(\vec{r}\cdot\vec{v})^2/v^2}} \Biggr] - B v r \vec{v},
\end{equation} 
which can be decomposed into radial and tangential components (c.f. eq.~\ref{eq:STW}) as
\begin{align}\label{eq:acc_comp}
S &= A v_r \left(\dfrac{v}{v_t} - \dfrac{v_r^2}{v v_t}\right) - B  v r  v_r, \\
T &= -A \dfrac{v_r^2}{v} - B v r  v_t,
\end{align}
where $A$ and $B$ are functions of $(a,e)$.  
Finally, we substitute $S$ and $T$ in eqs.~\ref{eq:a_variation} and~\ref{eq:e_variation}, and we tune the fitting functions $A$ and $B$ so that the orbit averages (eqs.~\ref{eq:a_av} and~\ref{eq:e_av}) match the results obtained by scattering experiments  
(eqs.~\ref{eq:a_evolution} and~\ref{eq:e_evolution}). 
If we choose the form 
\begin{align}\label{eq:const}
A(a,e) &= \dfrac{G\rho H K}{\sigma}\dfrac{2 a}{e(1-e^2)^{\beta_1}(1-e^5)^{\beta_2}}, \\
B(a,e) &= \dfrac{G\rho H}{2\sigma}\sqrt{\dfrac{a}{G M}}\Biggl[\dfrac{(1-e^{\beta_3})^{\beta_4}}{(1-e^{\beta_5})^{\beta_6}}  \Biggr],
\end{align}
where $\beta_1=0.38$, $\beta_2=0.055$, $\beta_3=8.036$, $\beta_4=0.148$, $\beta_5=1.90$, and $\beta_6=0.22$, we obtain a fairly good agreement with the results of three-body scattering experiments presented in \citet{Sesana2006} (see fig.~\ref{fig:hardening}). 

\subsection{Stellar bulge}
The hardening process described in the previous section depends upon the 
density profile of the stellar bulge hosting the MBHs. We describe the distribution of stars as a Hernquist profile \citep{Hernquist1990} with a central core:
\begin{equation}\label{eq:density}
 \rho(r) = 
  \begin{cases} 
   \dfrac{M_*}{2\pi} \dfrac{r_0}{r(r+r_0)^3}& \text{if } r > r_c, \\
   \ \\
   \rho(r_c)\left(r/r_c\right)^{-1/2}       & \text{if } r \leq r_c, 
  \end{cases}
\end{equation}
where $M_*$ and $r_0$ can be consistently determined by observational scaling relations \citep[e.g.,][]{Kormendy2013}, as described in \citet{Sesana2015}.
The inner core is modelled as a shallower power-law with index $-1/2$ \citep{2012ApJ...749..147K}, aimed at mimicking the erosion of the central region of the bulge by the now-stalled inner MBHB \citep{Ebisuzaki91, VolonteriCores2003, Antonini2015, 2015ApJ...806L...8A}. Indeed, during its alleged hardening phase, the inner MBH ejects stars via the slingshot mechanism, hence producing a {\it mass deficit} in the stellar distribution that can be quantified as \citep{Merrit2013book,Antonini2015,2015ApJ...806L...8A}
\begin{equation}\label{eq:mass_def}
\Delta M = M \left[0.7 q^{0.2} + 0.5\ln\left(0.178 \dfrac{c}{\sigma}\dfrac{q^{4/5}}{(1+q)^{3/5}}\right)\right],
\end{equation}
where here $M$ is the mass of the inner binary, and $q=m_2/m_1\leq 1$ the binary's mass ratio. The core radius $r_c$ can be easily obtained by imposing that $\Delta M$ equals the mass difference, within $r_c$, between the original Hernquist and $r^{-1/2}$ profiles .  

The bulge mass $M_b$ is given by integration of eq.~\ref{eq:density}, i.e., 
\begin{equation}\label{eq:stellar_mass}
 M_b(r) = 
  \begin{cases} 
   M_* r_0\biggl[\dfrac{4 r_c^2}{5(r_c+r_0)^3} \\
    + \dfrac{(r-r_c)(2r_c r + r_0(r_c+r))}{(r_c+r_0)^2(r+r_0)^2}\biggr] & \text{if } r > r_c \\
   \ \\
   \dfrac{4 M_* r^{5/2} r_0}{5 r_c^{1/2}(r_c+r_0)^3}       & \text{if } r \leq r_c.
  \end{cases}
\end{equation}
The core profile is then used to compute the hardening phase of the outer MBHB as detailed in the previous subsection, and to introduce a fixed analytical spherically symmetric potential in the equations of motion of the triplet, whose net effect is a Newtonian orbital precession with sign {\it opposite} to that induced by PN terms. 

Note that also during this phase stars will be ejected from the bulge, eroding the density profile and hence slowing the hardening of the outer binary. On the other hand, it is conceivable that a similar amount of mass in stars is brought in by the intruder, so the net effect is difficult  
to asses.  We therefore assume the outer binary to evolve in the full loss-cone limit, with the stellar distribution given 
by eq.~\ref{eq:density}.

\subsection{Stellar dynamical friction}
As a last effect acting already on kpc scale, we include the dynamical friction on the intruder of mass $m$ in its way to the bulge center. We adopt the simple Chandrasekhar's formula \citep{Chandrasekhar1943}, i.e., 
\begin{equation}\label{eq:dynamical_friction}
\dot{\vec{v}}_{\rm df} = -4\pi G \rho m \ln \Lambda \left[ {\rm erf}(X) - \dfrac{2 X e^{-X^2} }{\sqrt{\pi}}\right]\dfrac{\vec{v}}{v^3}
\end{equation}
where here $v$ is the velocity of the intruder, $\ln\Lambda$ is the Coulomb logarithm and $X = v/(\sqrt{2}\sigma)$. Following \citet{Hoffman2007}, we adopt
\begin{equation}\label{eq:Coulomb}
\ln\Lambda = {\rm max}\left\{ \ln\left( \dfrac{r(\sigma^2+v^2)}{G m} \right), 1 \right\},
\end{equation}
and
\begin{equation}\label{eq:density_dyn_fric}
\rho = {\rm min}\left\{ \rho(r), \rho(r_{\rm inf}) \right\},
\end{equation}
where $r_{\rm inf}$ is the binary influence radius. Dynamical friction is typically turned off in the code as soon as the intruder binds to the inner binary, when it becomes sub-dominant compared to the gravitational slingshot of background stars.

\begin{figure*}
 \begin{minipage}[c]{0.47\textwidth}
   \centering
   \includegraphics[scale=0.4]{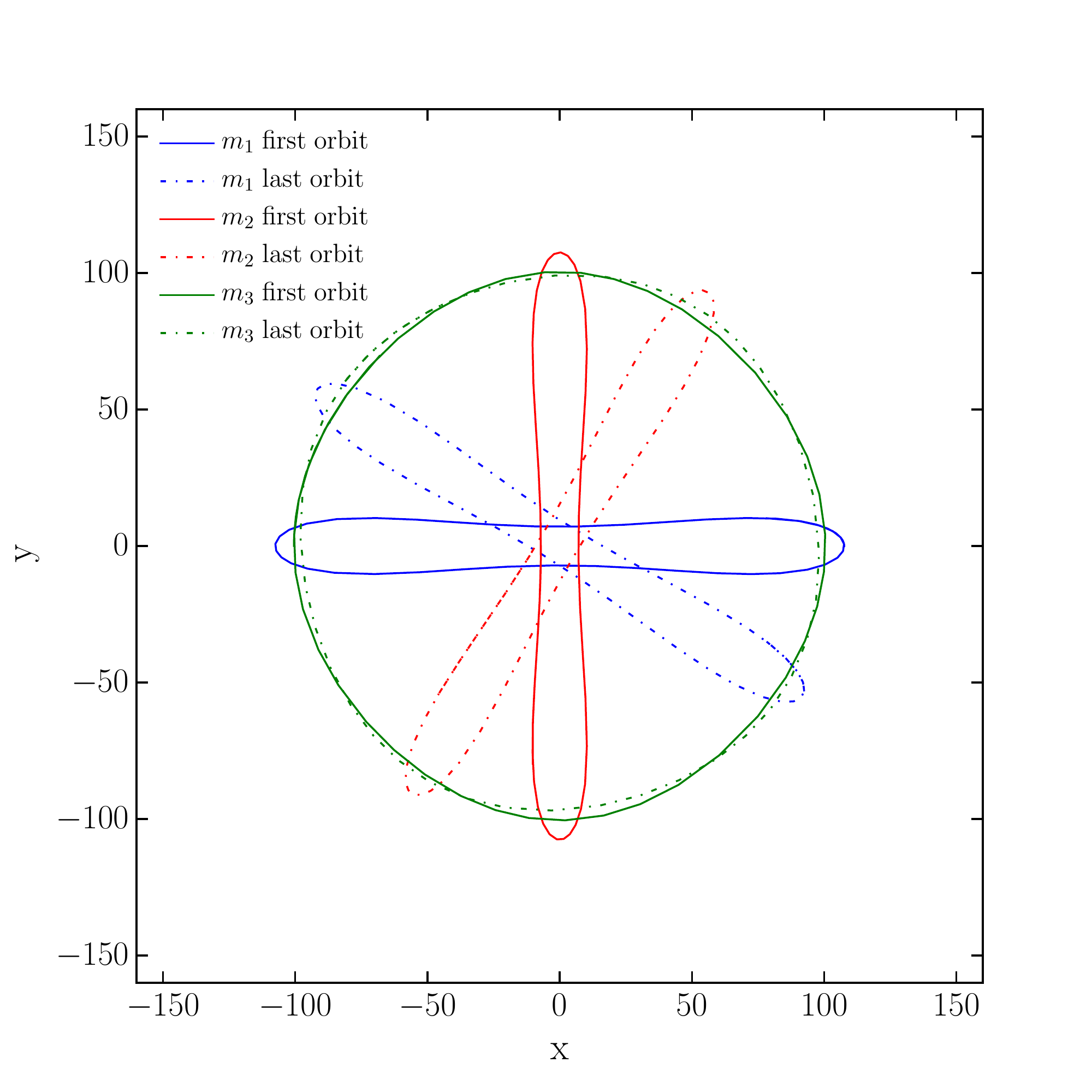}
 \end{minipage}
 \ \hspace{2mm} \hspace{3mm} \
 \begin{minipage}[c]{0.47\textwidth}
  \centering
   \includegraphics[scale=0.4]{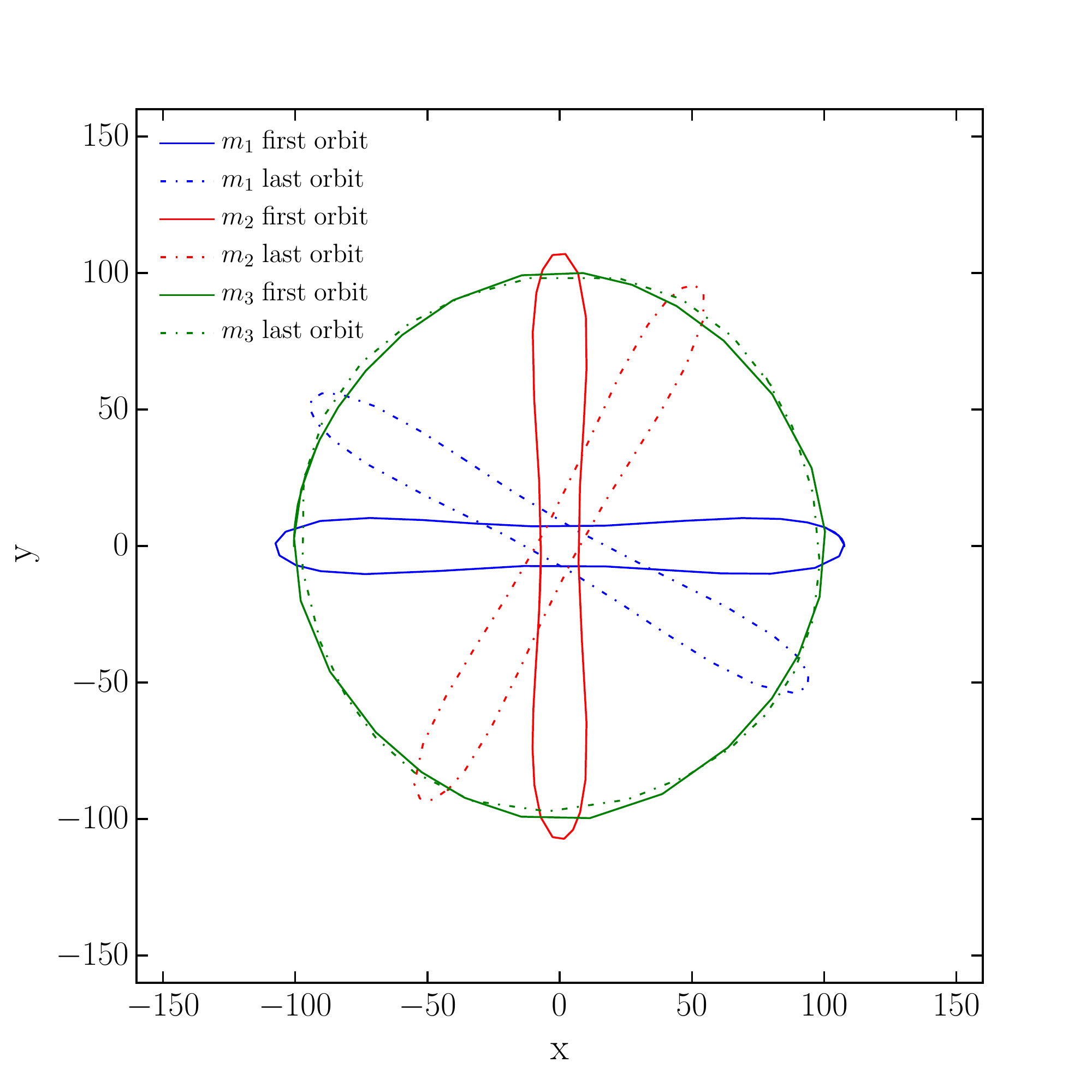}
 \end{minipage}
 \caption{Test of energy and angular momentum conservation for the Henon's Criss Cross configuration. Purely Newtonian dynamics is stable 
  against numerical errors for at least $10,000$ orbits. Integrations in quadruple (left panel) or double precision (right panel) do not show significant 
  differences. The apparent orbital precession has to be ascribed to round-off errors in the initial conditions.} 
  \label{fig:Newoniantest}
\end{figure*}

\section{Code implementation and tests}
We approached the computational problem from the most straightforward side, i.e., we employed a direct three-body integrator without any regularisation scheme to control round-off errors, rather, we selected a customised numerical precision suited for our goals.\footnote{We employed a direct three-body integrator since the secular equations of motion in some cases lead to inaccurate results, in particular when the eccentricities are very high \citep{Antonini2014}. Note that recently \citet{Luo2016} proposed a correction to the secular equations in order to recover the results of direct integration.}
Our numerical scheme directly integrates the three-body equations allowing for velocity-dependant forces (such as the PN dissipative terms and the binary hardening induced by the stellar background discussed in the previous section). 
The code leverages on a C++ implementation of the Bulirsch-Stoer (BS) method \citep{Bulirsch1966, NR} based on the Modified Midpoint algorithm and on the Richardson extrapolation \citep{Richardson1911}. The BS scheme advances the solution of a system of ordinary differential equations by ``macroscopic'' steps, i.e., the single step actually consists of many sub-steps of the Modified Midpoint method (i.e., the effective integrator scheme), which are then extrapolated to zero stepsize by the Richardson technique based on the Neville's algorithm \citep{NR}. The extrapolation, together with the dimension of the steps, provides a scheme to obtain high accuracy with minimised computational efforts. 

\subsection{Test of Newtonian dynamics}
In order to validate the code we performed some standard tests. 
We first tested the energy and angular momentum conservation at Newtonian order by means of a well known configuration for a three-body system, the so-called Henon's Criss-Cross \citep{Henon1976, Moore2006}. 
We evolved the system for nearly $10,000$ orbits, and checked that both energy and angular momentum are conserved at a level of one part in $10^{13}$. In fig.~\ref{fig:Newoniantest} we plot the first and last orbit of each of the three equal mass bodies of the test, comparing the results of 
integration in quadruple (left panel) and double (right panel) precision, and found no significant differences in the two runs. We will return later on this point. Note that the apparent orbital precession has to be ascribed to round-off errors in setting the appropriate initial conditions.

\subsection{Tests of PN dynamics}
We then proceeded to test our code against PN dynamics by performing some of the trial runs performed by \citet{Mikkola2008} 
using the ARCAHIN code. ARCHAIN, which includes non-dissipative 1PN, 2PN and dissipative 2.5PN corrections, employs a regularised chain structure and the time-transformed leapfrog scheme to accurately integrate the motions of arbitrarily close binaries with arbitrarily small mass-ratios. Recently, \citet{Antonini2016} used ARCHAIN to compute the evolution of hierarchical triplets formed in dense globular clusters. 

\begin{figure}
   \includegraphics[width=0.47\textwidth]{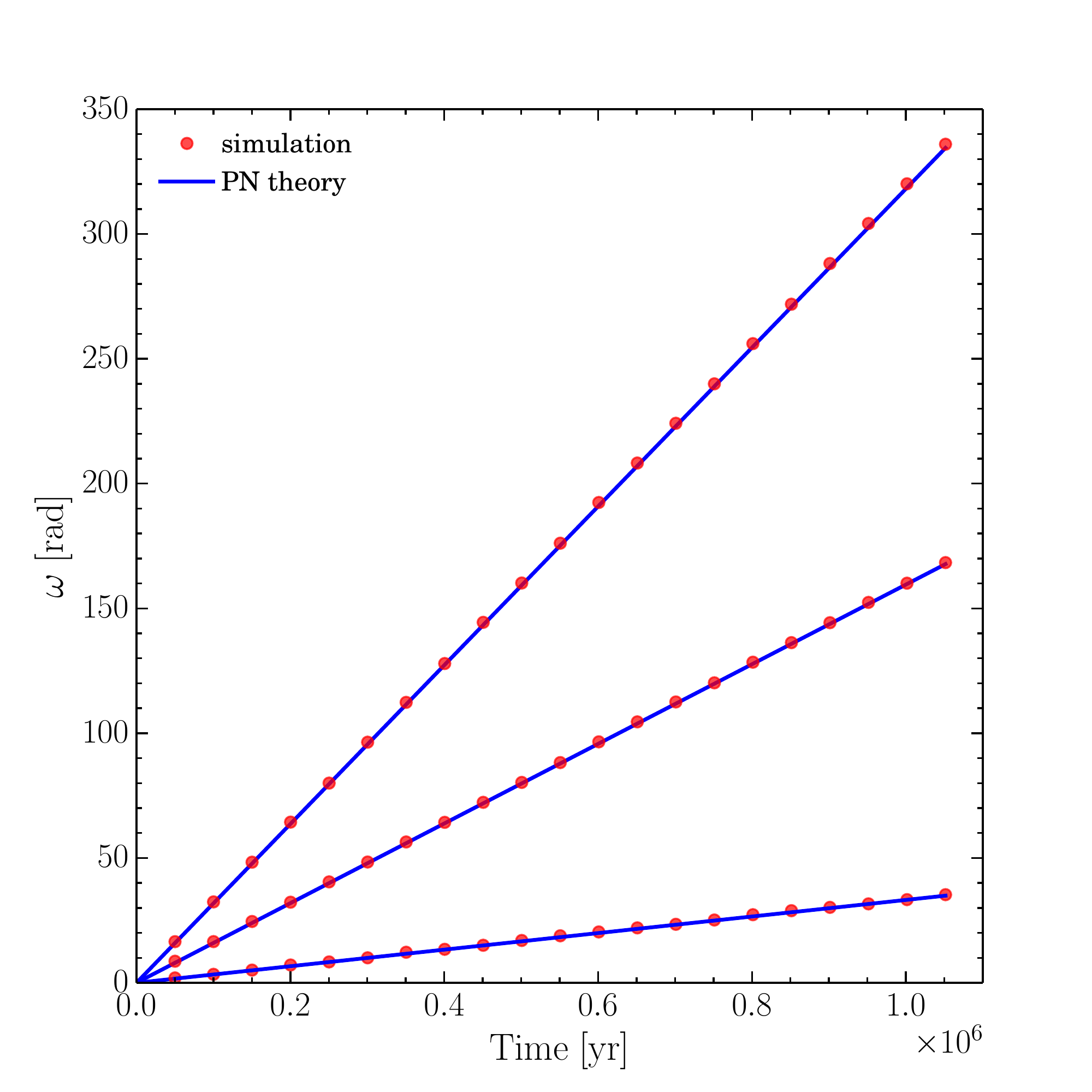}
   \caption{The periastron $\omega$ is shown against time for a two-body system consisting of a MBH of mass $m_{\rm BH} = 3.5\times 10^6~\msun$ and a star with $m_\star = 10~\msun$ and semi-major axis $a=1$ mpc. Red dots show every 1,000 orbits the advancement of the pericenter computed with our code, while the blue lines are the 2PN theoretical prediction. Three eccentricity are considered, $e=0.9,0.98,0.99$, from bottom to top.}
\label{fig:mikkola2}
\end{figure}

We report on two tests analogous to those presented in \citet{Mikkola2008}.
The first one consisted in a two-body dynamics check, where a star of mass $m_\star=10~\msun$ orbits a MBH of mass $m_{\rm BH}=3.5\times 10^6\msun$, with semi-major axis $a =1$ mpc. We considered 3 different eccentricities ($e =0.9,0.98,0.99$), and checked the progression of the periastron $\Delta\omega$ determined by the relativistic precession. As shown in fig.~\ref{fig:mikkola2}, we obtain in all tested cases a good agreement with the 2PN theoretical prediction, i.e., 
\begin{equation}
\Delta \omega = \dfrac{6\pi G M}{a(1-e^2)c^2} + \dfrac{3(18+e^2)\pi G^2 M^2}{2a^2(1-e^2)^2c^4},
\end{equation}
where $M=m_\star+m_{\rm BH}$.  

\begin{figure}
   \includegraphics[width=0.47\textwidth]{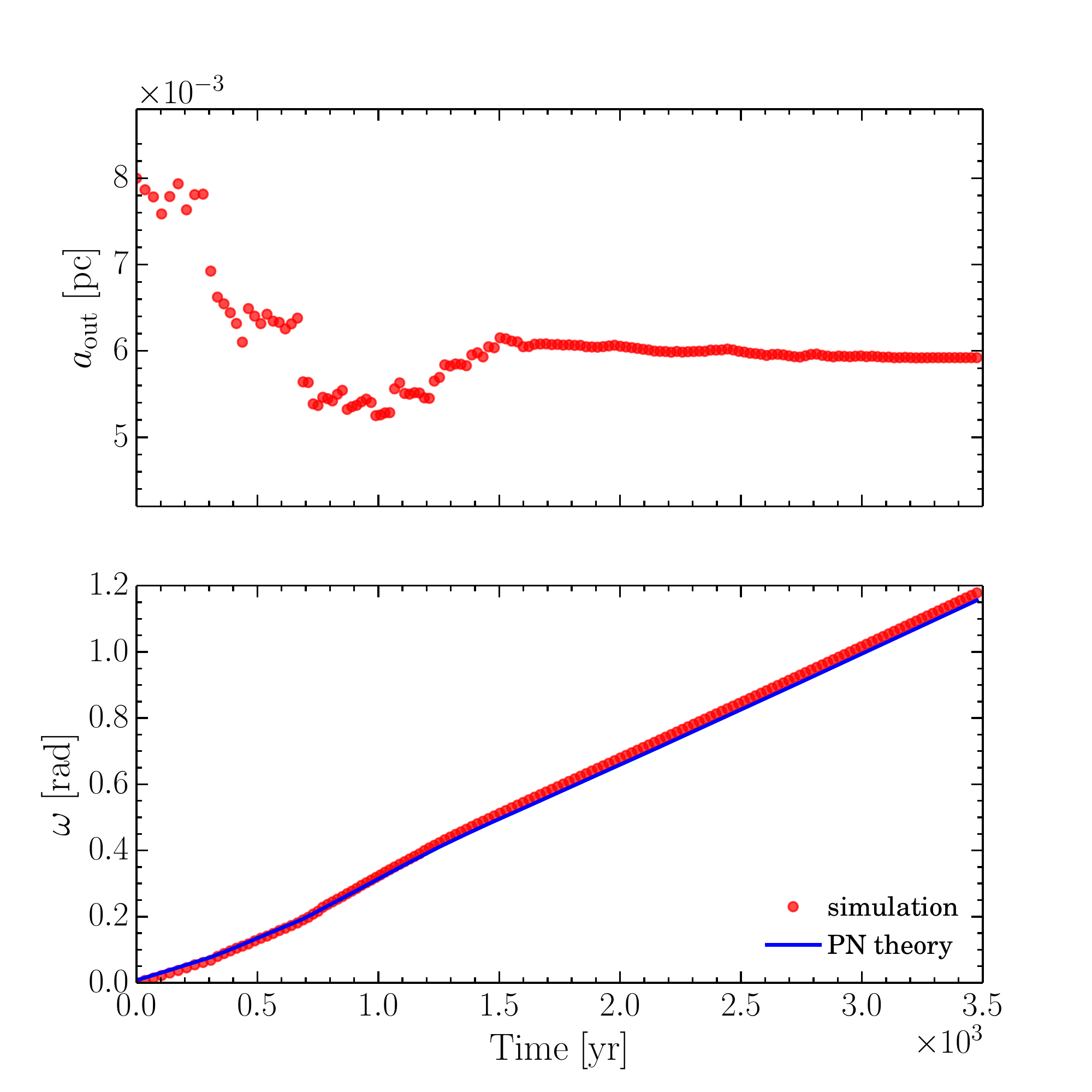}
  \caption{A three-body system with $m_1= 3.5\times 10^6~\msun$, $m_2 = 3.5\times 10^3~\msun$, and $m_3= 10~\msun$ is considered. 
The time evolution of the semi-major axis $a_{\rm out}$ (upper panel) and the argument of the pericenter $\omega$ of $m_3$ (lower panel) 
are shown as red dots. The blue line in the lower panel represents the 2PN theoretical predictions accounting for the time variations of $a_{\rm out}$ and $e_{\rm out}$.}  
 \label{fig:mikkola3}
\end{figure}

As a second test we analysed a three-body case, in which a star of mass $m_3 = 10\msun$ interacts with a MBHB, formed by a MBH of mass $m_1= 3.5\times 10^6\msun$, and an intermediate mass BH with $m_2 = 3.5\times 10^3\msun$. The MBHB has semi-major axis $a_{\rm in}= 0.1$ mpc and eccentricity $e_{\rm in}= 0.9$, while the star is placed on an orbit with $a_{\rm out}= 8$ mpc and $e_{\rm out}= 0.974$. In fig.~\ref{fig:mikkola3}, upper panel, we show $a_{\rm out}$ as a function of time. Initially the star experiences close encounters with the MBHB, as apparent from the ``noisy" pattern of $a_{\rm out}$. After $\simeq 1,500$ yrs 
the separation of the MBHs has greatly reduced because of GW emission, the star effectively ``sees" an almost central potential at this stage, and 
its orbital separation stabilises. In fig.~\ref{fig:mikkola3}, lower panel, the argument of the star pericenter $\omega$ is compared to 2PN theoretical predictions accounting for the time variations of $a_{\rm out}$ and $e_{\rm out}$. 
Overall our results are in close agreement with those reported in \citet{Mikkola2008}, although some minor numerical differences do exist, most 
probably due to slightly different initial conditions \citep[e.g., the MBHB initial phase was not reported in][while we initially placed both the MBHB and the star at the respective apocenter]{Mikkola2008}.

\begin{figure*}
 \begin{minipage}[b]{0.47\textwidth}
   \centering
   \includegraphics[scale=0.4]{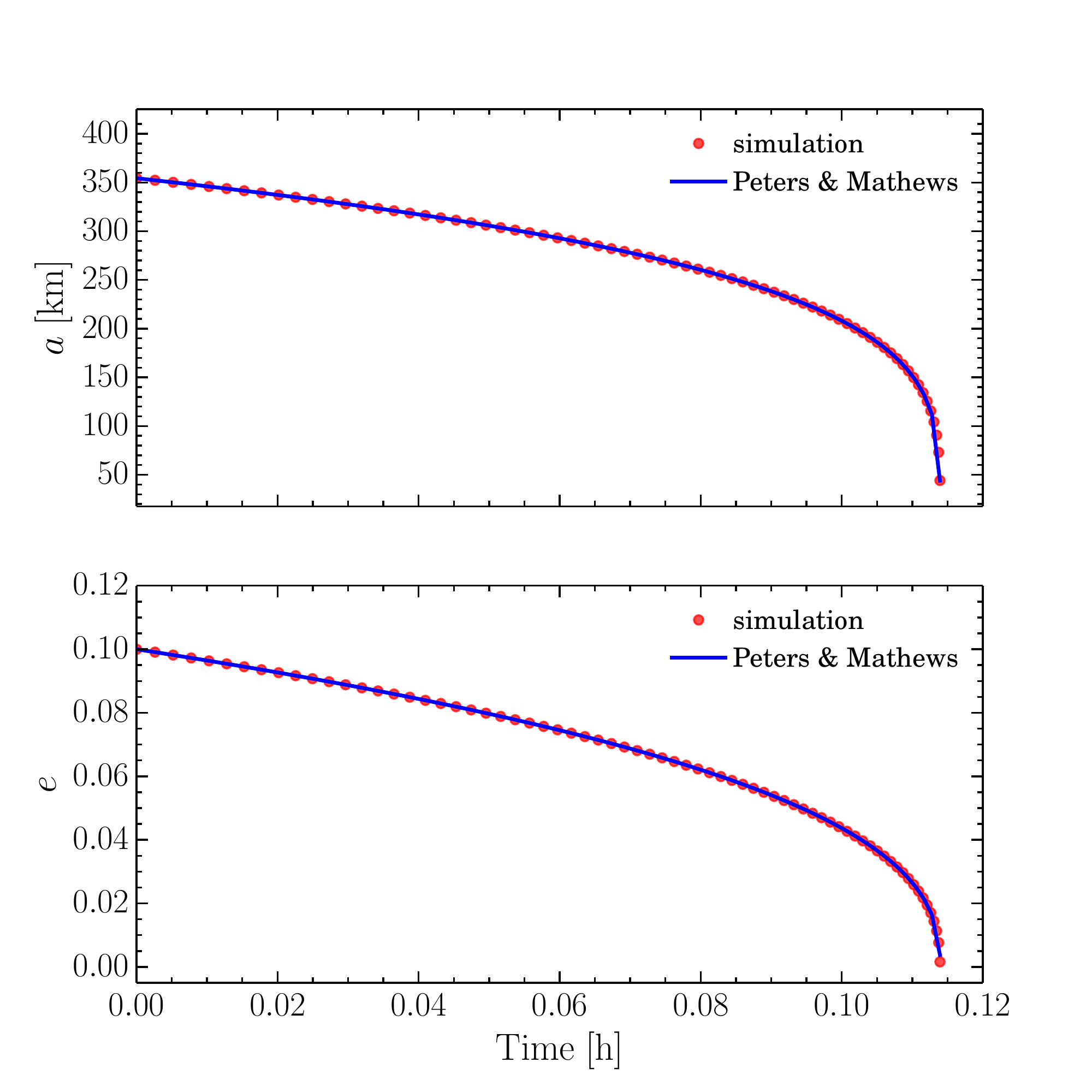}
 \end{minipage}
 \ \hspace{2mm} \
 \begin{minipage}[b]{0.47\textwidth}
  \centering
   \includegraphics[scale=0.4]{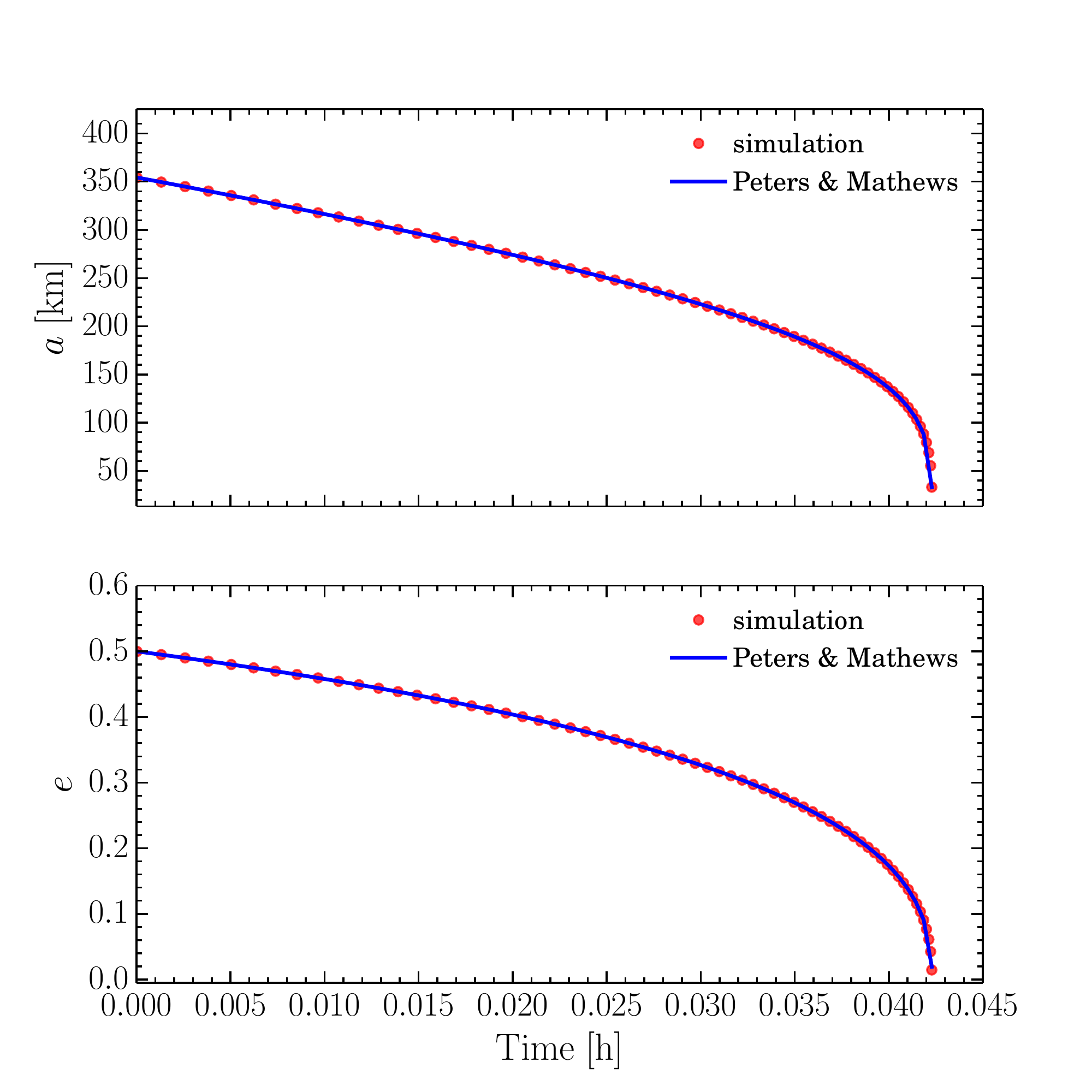}
 \end{minipage}
 \caption{The orbital evolution of a binary system with masses $2m_2=m_1 = 1~\msun$ including 2.5PN terms only (red dots) is compared to the evolution computed by employing the orbital averaged GW losses from \citet{Peters1963} (blue lines). The upper panels show the binary semi-major axis, the lower panels the binary eccentricity as a function of time (in hours). {\it Left}: the initial binary eccentricity is set to $e=0.1$. {\it Right}: the initial eccentricity is set 
 to $e=0.5$.}  
 \label{fig:peters}
\end{figure*}

We then tested the implementation of the dissipative 2.5PN term, comparing our results to the numerical integration of the orbit-averaged equation of \citet{Peters1963} for the time derivative of the semi-major axis and eccentricity, i.e.,
\begin{align}\label{eq:dadtGW}
\dot a&= -\dfrac{64 G^3}{5 c^5}\dfrac{m_1 m_2}{a^3(1-e^2)^{7/2}}\left(1+\dfrac{73}{24}e^2+\dfrac{37}{96}e^4\right)\\
\dot e &= -\dfrac{304 G^3}{15 c^5}\dfrac{m_1 m_2}{a^4(1-e^2)^{5/2}}\left(e+\dfrac{121}{304}e^3\right).\label{eq:dedtGW}
\end{align}
We selected a stellar-size binary \citep[see][]{Galaviz2011} with $m_1=1~\msun$, $m_2=m_1/2$, initial semi-major axis $a=160 G(m_1+m_2)/c^2$, and two different values of the initial eccentricity, $e=0.1$ and $e=0.5$. We switched off the 1PN and 2PN terms, as eqs.~\ref{eq:dadtGW} and \ref{eq:dedtGW} take into account only 2.5PN order corrections to the Newtonian dynamics. Fig.~\ref{fig:peters} shows the excellent agreement between simulations and analytical results. 

\subsection{Effects of numerical precision}
\begin{figure*}
 \begin{minipage}[b]{0.47\textwidth}
   \centering
   \includegraphics[scale=0.4]{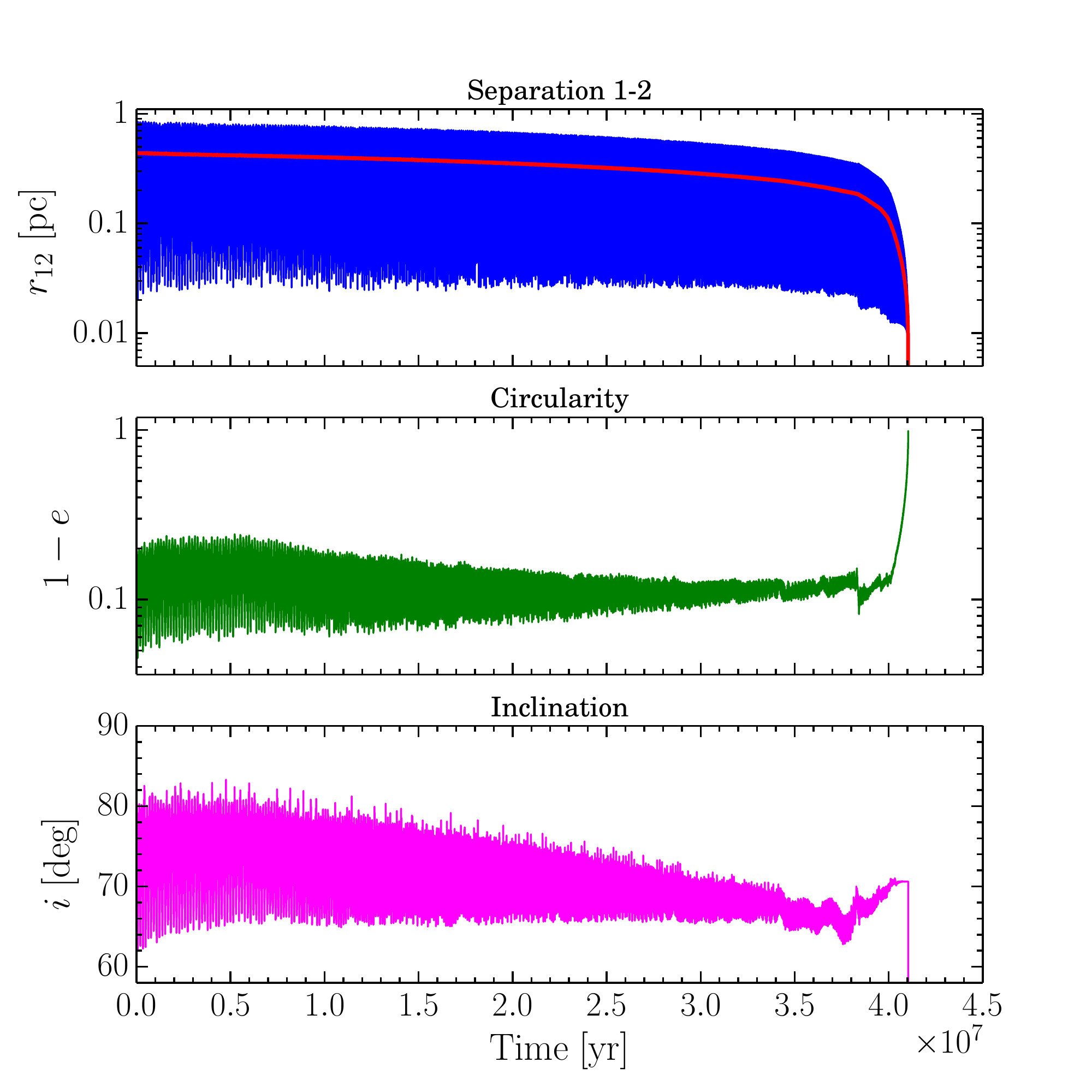}
 \end{minipage}
 \ \hspace{2mm} \
 \begin{minipage}[b]{0.47\textwidth}
  \centering
   \includegraphics[scale=0.4]{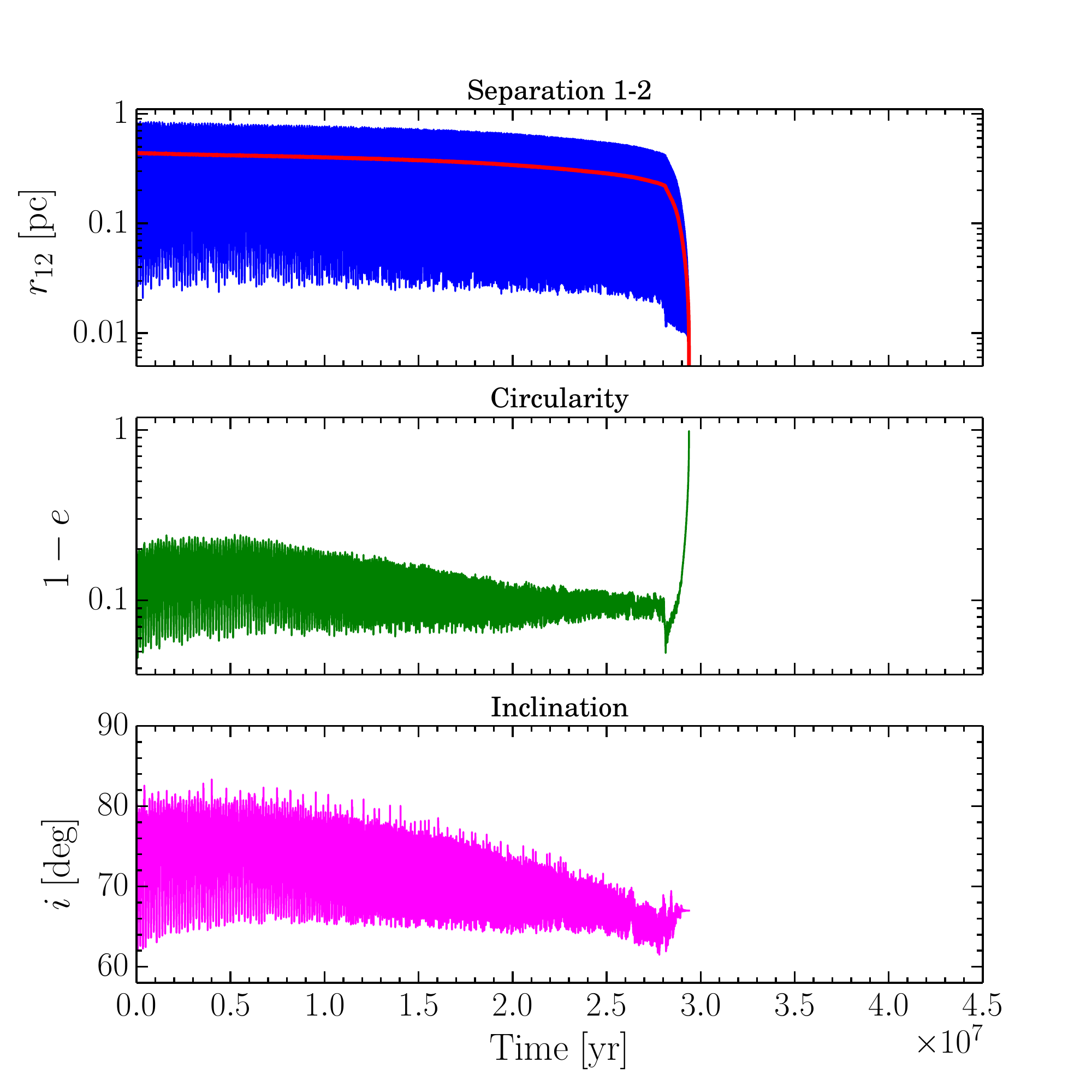}
 \end{minipage}
 \caption{The relative separation (upper panels), circularity (middle panels) and inclination (lower panels) are shown for 
the inner binary of a hierarchical triplet with $m_1=10^9\msun$, $m_2=3\times 10^8\msun$, $m_3=5\times 10^8\msun$, $a_{\rm out} = 4.43$pc, $e_{\rm out} = 0.5$, $a_{\rm in} = 0.44$pc, $e_{\rm in} = 0.8$, and $i = 80^\circ$. The solid red line is $a_{\rm in}$. {\it Left}: quadruple precision calculation. 
{\it Right}: double precision calculation.}   
 \label{fig:test_caos}
\end{figure*}
\begin{figure*}
 \begin{minipage}[b]{0.47\textwidth}
   \centering
   \includegraphics[scale=0.4]{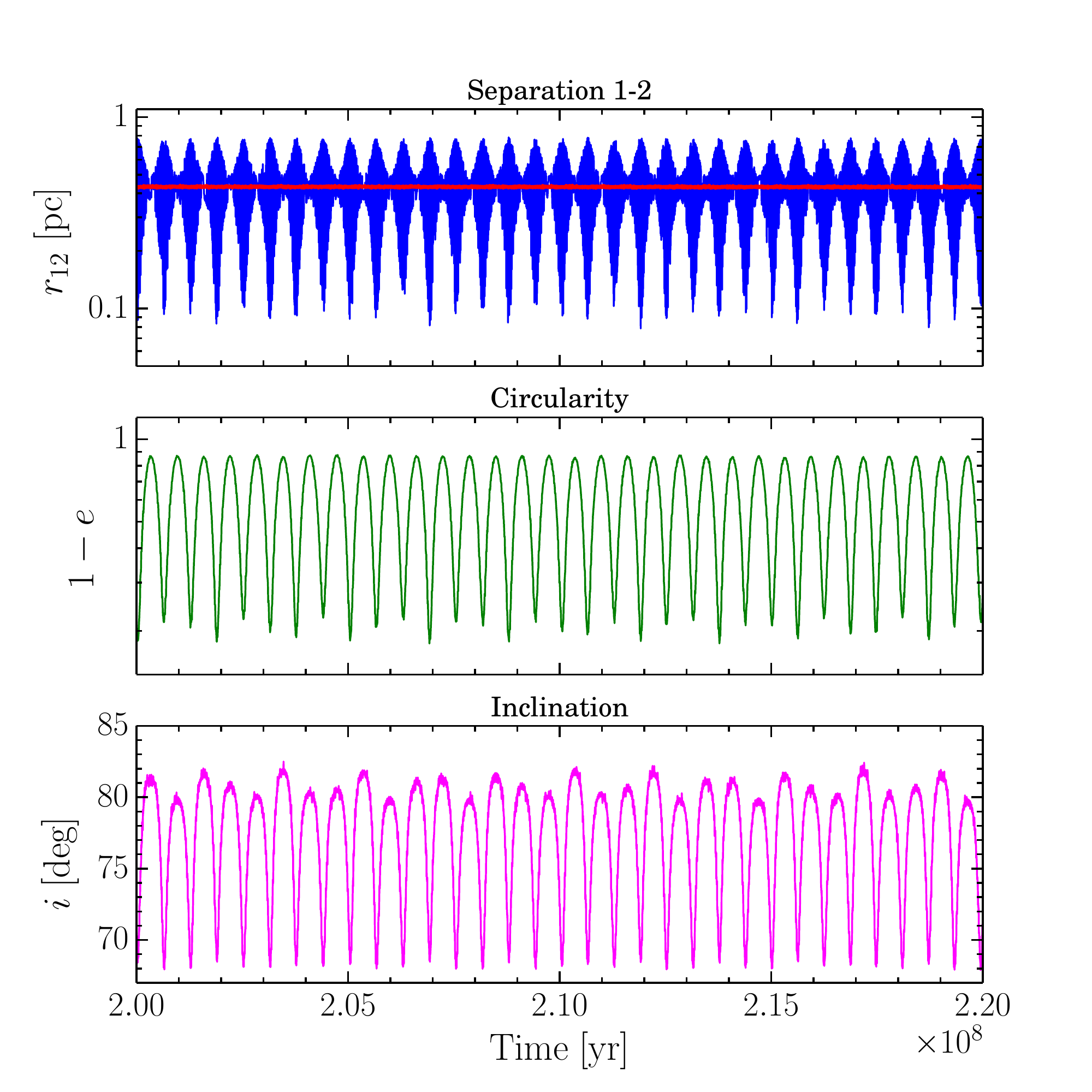}
 \end{minipage}
 \ \hspace{2mm} \
 \begin{minipage}[b]{0.47\textwidth}
  \centering
   \includegraphics[scale=0.4]{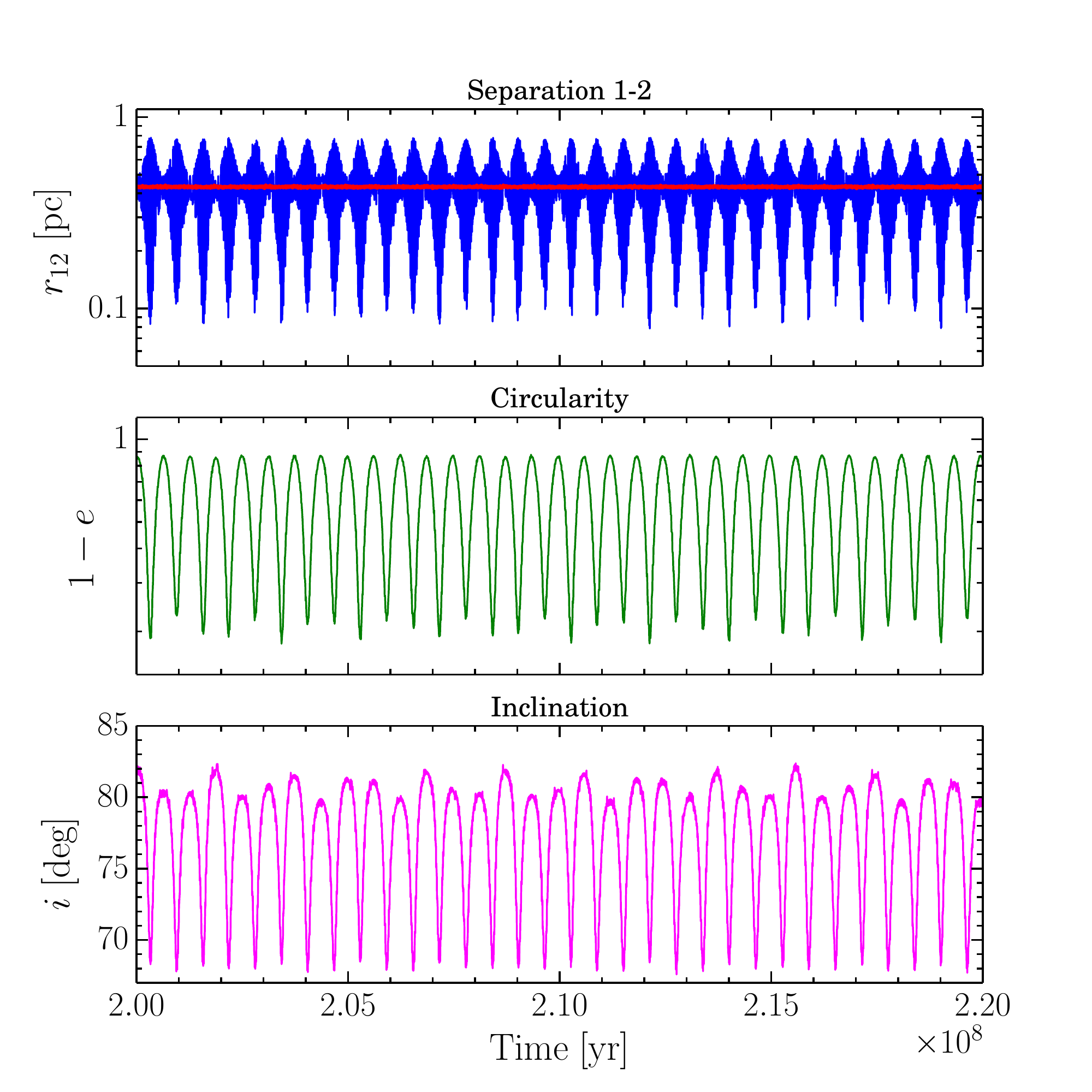}
 \end{minipage}
 \caption{Same as fig.~\ref{fig:test_caos} but assuming an initial $e_{\rm in}=0$. Note the different temporal scale compared to 
 fig.~\ref{fig:test_caos}.}
 \label{fig:test_caos2}
\end{figure*}

We return here on the impact that the adopted numerical precision has on our results. We have already seen that in the Henon's Criss-Cross test 
the double and quadruple precisions give essentially the same output. We performed a further detailed analysis of the issue comparing simulations of the full dynamics of MBH triplets (i.e., all terms up to 2.5PN were considered). We found that the intrinsic chaotic nature of the three-body problem makes results quantitatively dependent upon the  chosen numerical precision.\footnote{This is known as the ``shadowing'' property of numerical solutions to deterministic chaotic systems. 
Indeed, for these systems, the details of a numerical solution are highly dependent on the round-off errors, but the calculated solution
is very close to \textit{some} trajectory of the system, i.e., it may not correspond exactly to the desired trajectory, but
to another possible trajectory of the system (namely, one with slightly different initial conditions). See, e.g., \citet{shadowing}.}
However, the qualitative behaviour of the simulated triplets appear to be fairly robust against round-off errors. This is shown in fig.~\ref{fig:test_caos}, where we plot the relative separation (upper panels), circularity (middle panels) and inclination (lower panels) of the inner binary. Note that single orbits are not recognisable on this time scale. The red line represents the semi-major axis of the inner binary. Results in quadruple (double) precision for a triple system with: $m_1=10^9\msun$, $m_2=3\times 10^8\msun$, $m_3=5\times 10^8\msun$, $a_{\rm out} = 4.43$pc, $e_{\rm out} = 0.5$, $a_{\rm in} = 0.44$pc, $e_{\rm in} = 0.8$, and $i = 80^\circ$ are shown in the left (right) panels. In both cases the inner binary is bound to coalesce in few times $10^7$ yrs, the precise timescale depending upon the adopted numerical precision. In fig.~\ref{fig:test_caos2} we report the very same quantities for a similar triplet with initial $e_{\rm in}=0$. In order to highlight the Kozai-Lidov oscillations experienced by the inner binary, we show a time zoom of the orbital evolution. Note that on this scale the only clear difference between quadruple and double precision integration is a slight temporal shift of the whole evolution. 

We conclude our analysis of the effects of numerical precision by pointing out that quadruple precision typically takes, in terms of computer time, at least a factor $\gsim 10$ longer than double precision for the same set of parameters. Given this fact, the similar behaviour we witnessed in the test cases, and the fact that our final goal is a detailed survey of the parameter space of MBH triplets in a cosmological context, we decided to restrict our analysis to simulations in double numerical precision. 

\section{Dynamics of MBH triplets }

\subsection{Standard three-body dynamics}
\begin{figure*}
 \begin{minipage}[b]{0.47\textwidth}
   \centering
   \includegraphics[scale=0.4]{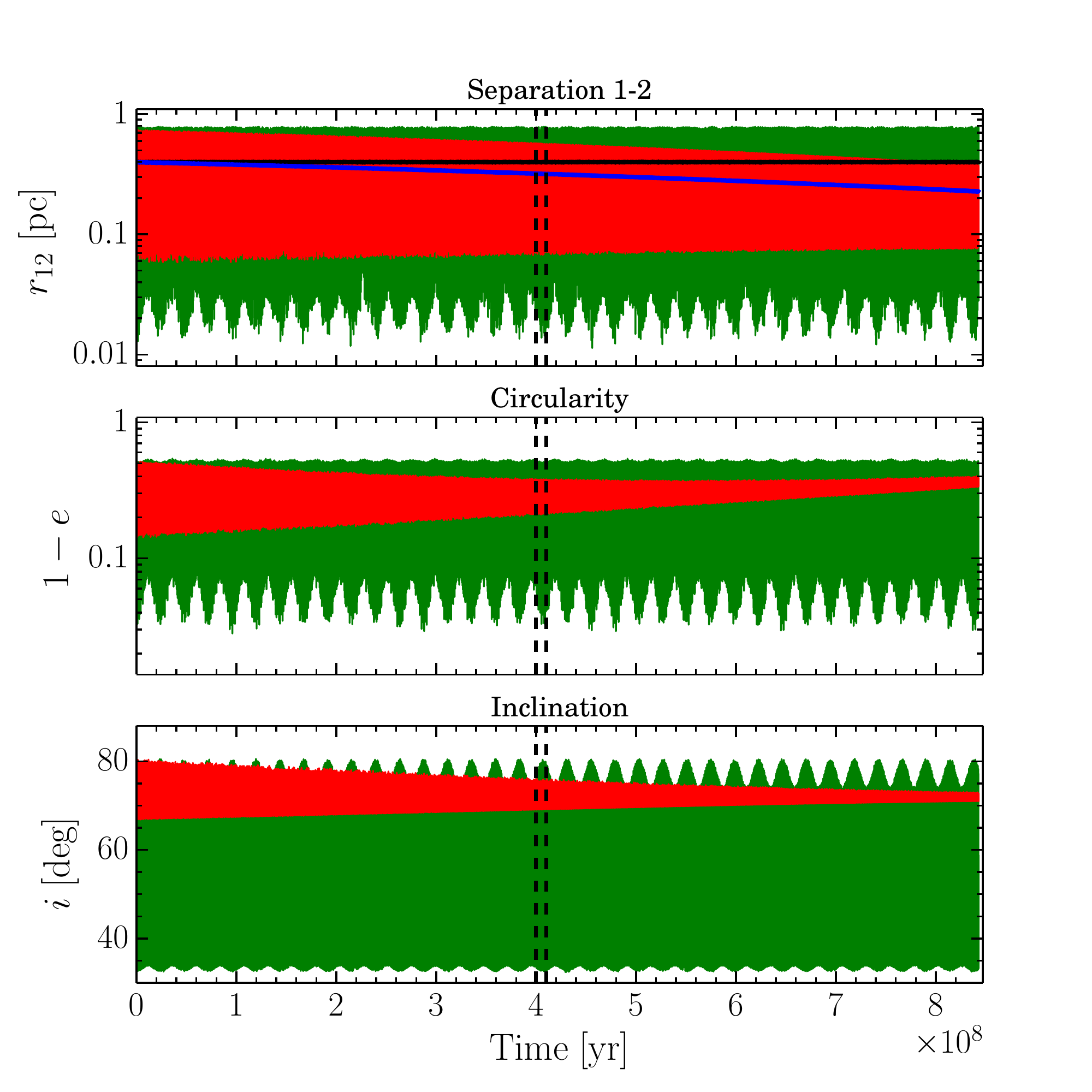}
 \end{minipage}
 \ \hspace{2mm} \
 \begin{minipage}[b]{0.47\textwidth}
  \centering
   \includegraphics[scale=0.4]{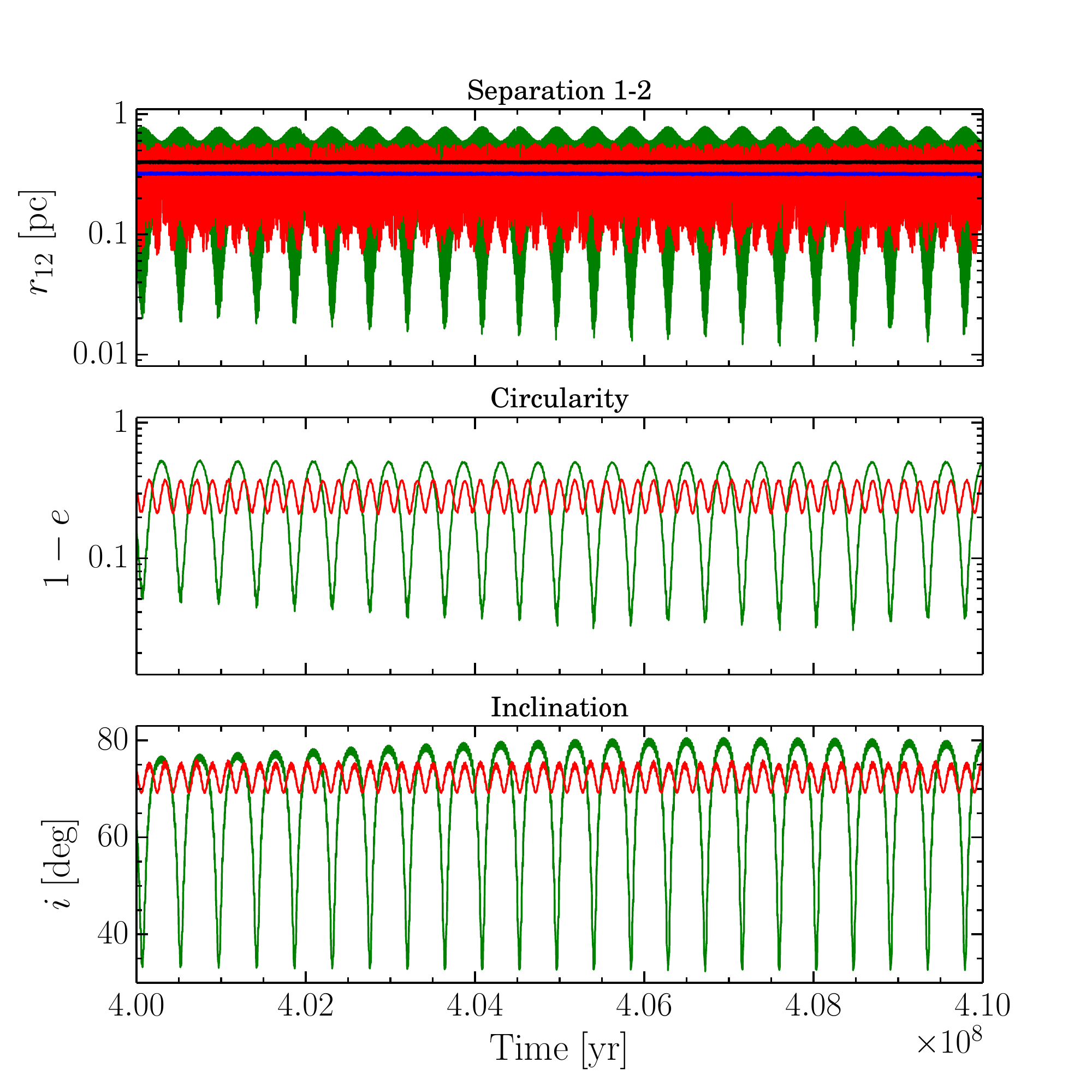}
 \end{minipage}
 \caption{A triplet with $m_1=10^9\msun$, $m_2=3\times 10^8\msun$, $m_3=5\times 10^8\msun$, $a_{\rm out} = 4$ pc, $e_{\rm out} = 0.5$, $a_{\rm in}=0.4$ pc, $e_{\rm in} = 0.3$, and $i = 80^{\circ}$ is considered. Relative separation (upper panels), the circularity (middle panels) and the inclination (lower panels) of the inner binary are plotted against time. Red colour refers to results from 2.5PN calculations, while the corresponding Newtonian values are shown in green. The blue (black) line represents the 2.5PN (Newtonian) value of the semi-major axis $a_{\rm in}$. {\it Left}: entire run. The vertical dashed lines frame the time interval zoomed in the right panels. {\it Right}: the 10 Myr time zoom.}   
 \label{fig:triplet1}
\end{figure*}
\begin{figure*}
 \begin{minipage}[b]{0.47\textwidth}
   \centering
   \includegraphics[scale=0.4]{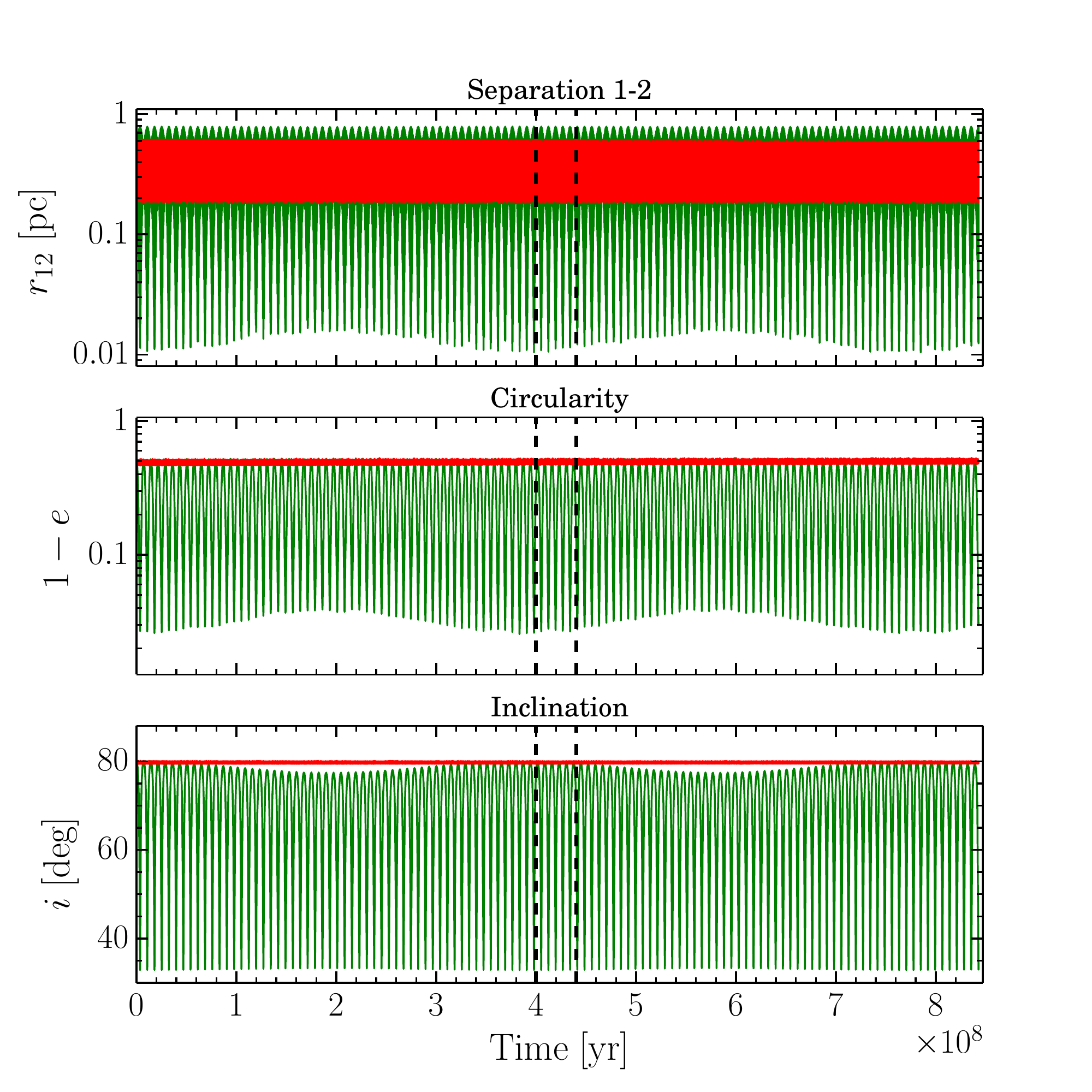}
 \end{minipage}
 \ \hspace{2mm} \
 \begin{minipage}[b]{0.47\textwidth}
  \centering
   \includegraphics[scale=0.4]{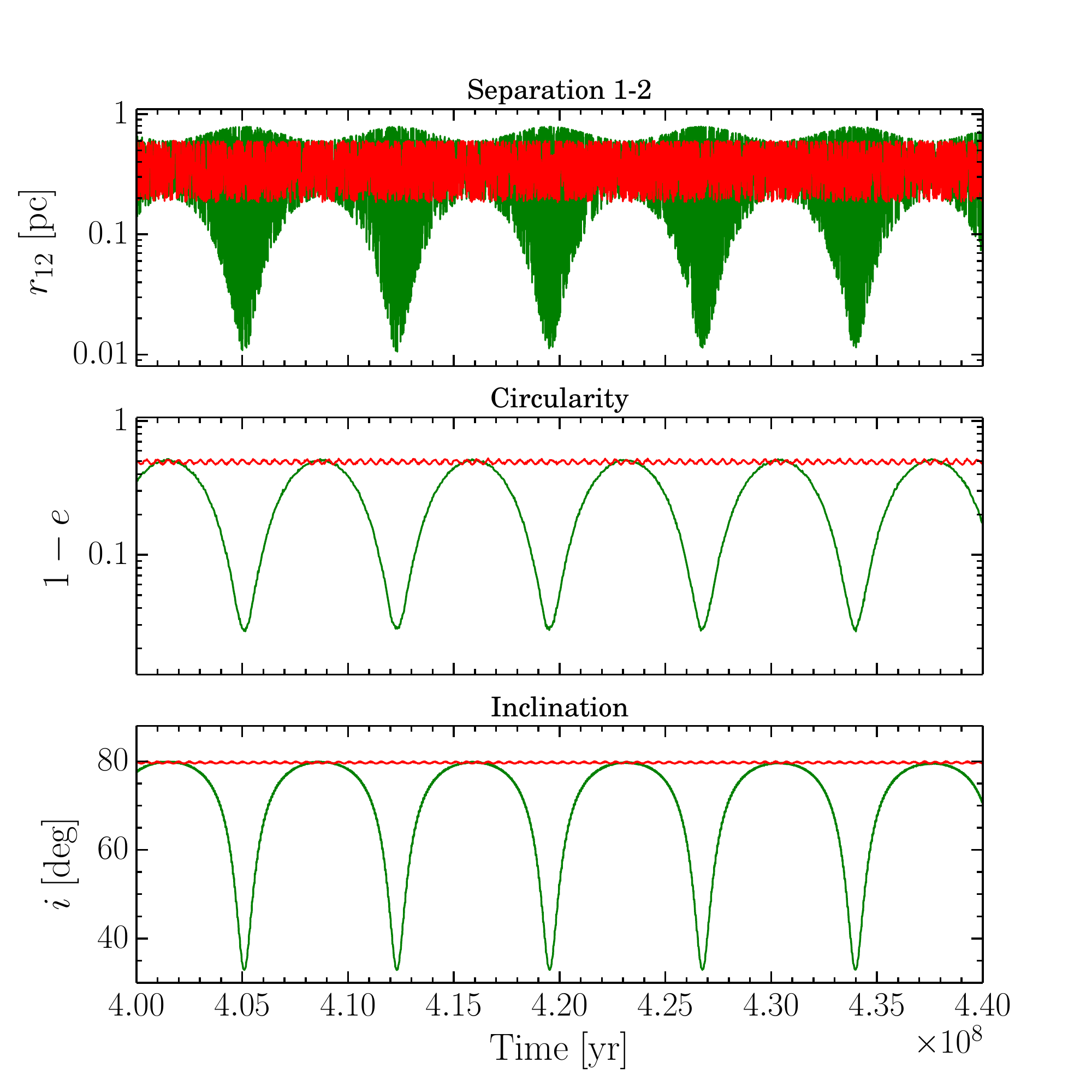}
 \end{minipage}
 \caption{Same as fig.~\ref{fig:triplet1} but for $a_{\rm out}=10$ pc. Here $a_{\rm in}$ is not plotted as it stays nearly constant even in the 2.5PN 
 calculations. {\it Left}: entire run. The vertical dashed lines frame the time interval zoomed in the right panels. {\it Right}: the 40 Myr time zoom.}  
 \label{fig:triplet2}
\end{figure*}

We start by considering the standard three-body problem, i.e., the interaction of three MBHs (including relativistic corrections 
up to order 2.5PN) without the inclusion of any external force due to, e.g., the stellar environment. We study the particular case of (initially) hierarchical triplets, 
i.e., three-body systems that can be modelled as two separate binaries: an inner, close one ($m_1$ and $m_2$) 
and a much wider one formed by the intruder $m_3$ and the center of mass of the former.  

A hierarchical triplet is prone to a peculiar dynamical phenomenon of purely Newtonian origin, known as the Kozai-Lidov (K-L) mechanism \citep{Kozai1962,Lidov1962}. If the relative inclination between the inner and outer binary is larger than a critical angle ($\simeq 39.23^\circ$), then 
a periodic exchange between the inclination and the eccentricity of the inner binary occurs on a time-scale 
\begin{equation}
t_{\rm KL} \sim \dfrac{a_{\rm out}^3(1-e_{\rm out}^2)^{3/2}\sqrt{m_1+m_2}}{G^{1/2} a_{\rm in}^{3/2}m_3 }\simeq 2\times10^6\,\,\,\, {\rm yrs},
\label{eq:TKozai}
\end{equation}
where the numerical  value reported above is for a MBHB with $m_1=m_2=m_3=10^8~\msun$, $a_{\rm in}=1$ pc, $a_{\rm out}=10$ pc, and $e_{\rm out}=0$.

The most important feature of the K-L mechanism is the excitement of periodic oscillations of $e_{\rm in}$ at the expenses of the relative inclination of the two binaries: $e_{\rm in}$ has a maximum when the relative inclination of the two binaries reaches its minimum value and viceversa. In this 
situation the emission of GWs is highly efficient, hence the K-L mechanism ultimately promotes the coalescence of the inner binary, possibly easing the last-parsec problem. Indeed, the oscillations of $e_{\rm in}$ can in principle reduce the coalescence time-scale by orders of magnitude. 

However, as already pointed out, the K-L mechanism is a purely Newtonian phenomenon, and the inclusion of relativistic corrections to Newtonian dynamics can have dramatic consequences on the K-L mechanism itself \citep[see][]{Blaes2002}. The very process relies on the libration of the pericenter argument of the inner binary. Since relativistic effects cause the precession of this orbital element, if the K-L timescale is longer than the relativistic precession timescale, the oscillations of $e_{\rm in}$  can be strongly inhibited compared to the purely Newtonian case. 

As an example, we consider a triplet with the following parameters: $m_1=10^9\msun$, $m_2=3\times 10^8\msun$, $m_3=5\times 10^8\msun$, $e_{\rm out} = 0.5$, $a_{\rm in} = 0.4$pc, $e_{\rm in} = 0.3$, and $i = 80^{\circ}$, and two different values of $a_{\rm out}$. 
In fig.~\ref{fig:triplet1} we compare the 2.5PN (red lines) and purely Newtonian (green lines) dynamical evolution of the system, assuming $a_{\rm out}=4$ pc. We plot the relative separation (upper panels), circularity (middle panels) and inclination (lower panels) of the inner binary.  
The left panels display the evolution for a time spanning almost 1 Gyr. 
The oscillations present in the Newtonian case are the modulation 
(ascribable to the octupole term) of the K-L oscillations (due instead to the quadrupole term). Note how the effect of GW emission manifests itself in the PN quantities, i.e., in a decrease of the semi-major axis (blue line) compared to the Newtonian case (black line), and in orbital circularisation. 
The K-L oscillations are resolved in the right panels, where a time zoom of 10 Myr highlights the significant damping of K-L oscillations when relativistic effects are included. 
\begin{figure}
\includegraphics[width=0.47\textwidth]{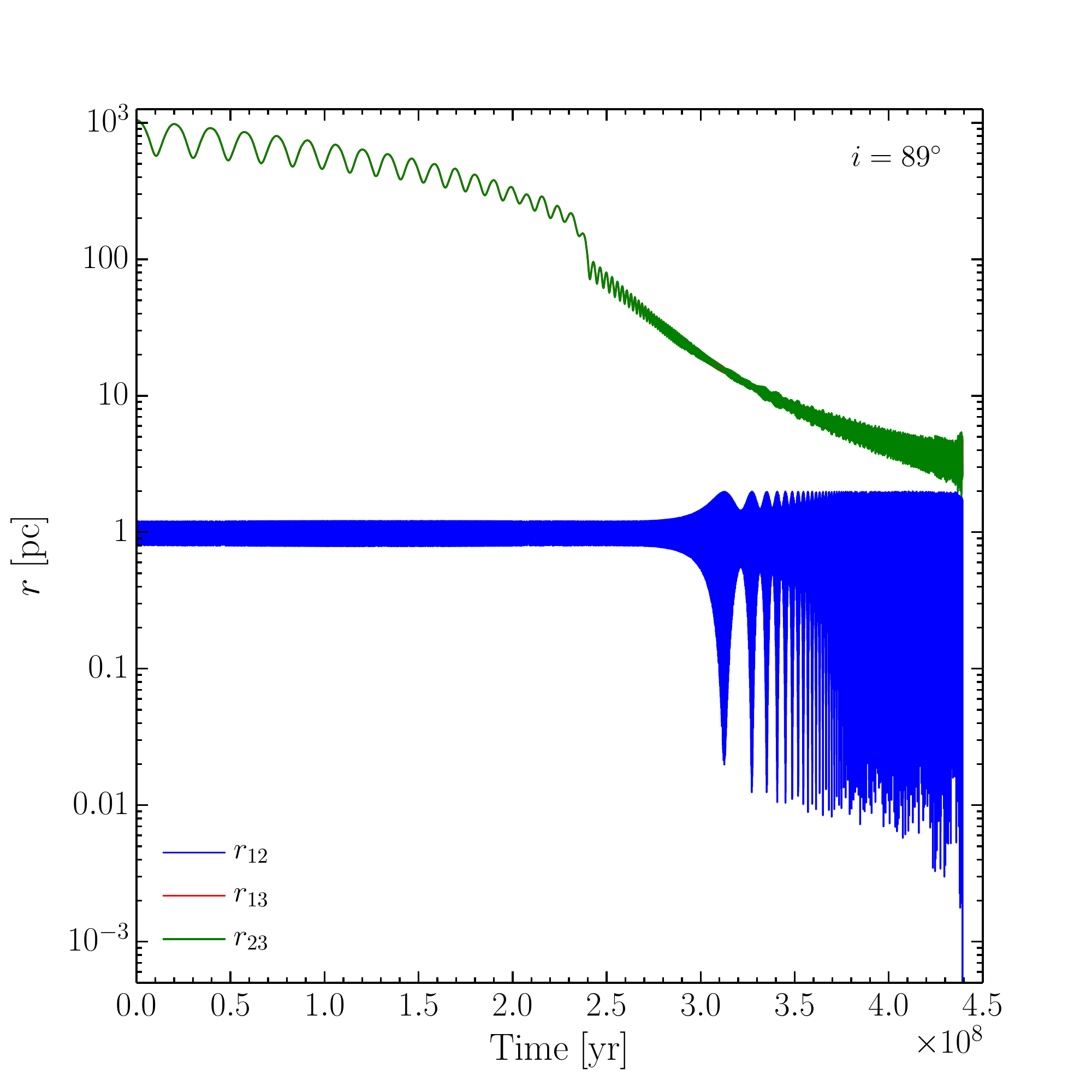}
\caption{Full evolution of a MBH triplet ($m_1=10^8\msun$, $m_2=3\times 10^7\msun$, $m_3=5\times 10^7\msun$, $a_{\rm in} = 1$pc, $e_{\rm in} = 0.2$, $i = 89^{\circ}$) in a stellar environment. The relative separation (in log scale) between $m_1$ and $m_2$ is shown in blue, that between $m_2$ and $m_3$ in 
green, and that (though not clearly visible, see text for details)  between $m_1$ and $m_3$ in red.} 
\label{fig:test_all_1}
\end{figure}
\begin{figure}
\includegraphics[width=0.47\textwidth]{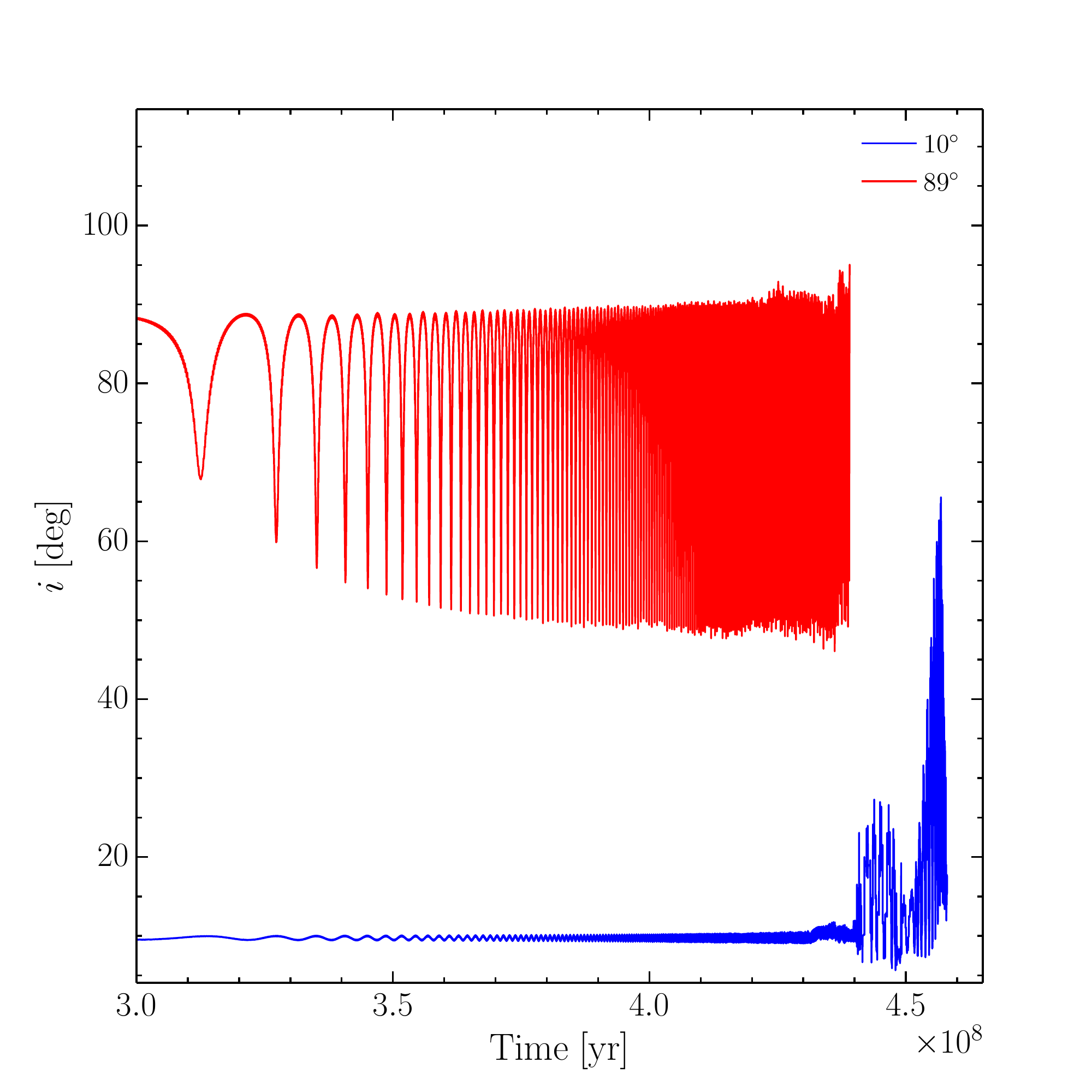}
\caption{Time evolution of relative inclination $i$ of the two binaries. Upper red curve is for the case of an initial value $i=89^\circ$, 
lower blue curve is for $i=10^\circ$.} 
\label{fig:inc_vs_t89}
\end{figure}

Fig.~\ref{fig:triplet2} shows the very same triplet, but assuming $a_{\rm out}=10$ pc. Since the K-L timescale has a strong dependence upon 
$a_{\rm out}$ (see eq.~\ref{eq:TKozai}), in this case relativistic precession is comparatively shorter, hence completely 
destroying the K-L oscillations.

\subsection{Three-body dynamics in stellar environments}

We finally analyse few examples of triplet dynamics including the effect of the stellar background, as discussed in \S 2. 
Our initial conditions consist of an inner MBHB ($m_1=10^8\msun$, $m_2=3\times10^7\msun$) stalled on an elliptical orbit ($e_{\rm in}=0.2$) at the center of a spherical stellar distribution (c.f. eqs.~\ref{eq:density}--\ref{eq:stellar_mass}), and a third, initially far MBH ($m_3=5\times 10^7\msun$) sinking in the global potential well because of dynamical friction. We initialise $m_3$ on an elliptical orbit ($e_{\rm out}=0.3$) at a distance from the center of the order of the bulge scale radius. The properties of the stellar distribution (see \S2.3) are determined  by 
the mass of the inner MBHB following the scaling relations of \citet{Kormendy2013} 
\citep[for full details see][]{Sesana2015}. Our choice of the inner binary mass then gives a stellar mass $M_\star=3.2\times 10^{10}\msun$ (the corresponding velocity dispersion is $\sigma=164$ km/s), a scale radius $r_0=1.2$ kpc, and a core radius $r_c=270$ pc. 

In the simulations, dynamical friction (\S2.4) operates on $m_3$ from the very beginning, while  
the hardening process (\S2.2) is activated when the intruder reaches the influence radius of the outer binary (i.e., 
for a hierarchical triplet, the radius containing a mass in stars twice the total mass of the three MBHs). As the triplet evolution proceeds, the hardening process on the outer binary becomes more important than dynamical friction on $m_3$. Then, we turn off effect of dynamical friction as soon as the intruder effectively binds to the inner binary, hence forming a genuine, hierarchical bound triplet.
\footnote{Note that a bound binary is 
not necessarily hard \citep[i.e., dynamical friction operates  on a light intruder well within the influence radius of the inner binary, see e.g.,][]
{Antonini2012}.  
We therefore run test simulations allowing the dynamical friction to continuously operate after the formation of a bound triplet, and we 
checked that the overall evolution of the system is not qualitatively different compared to our standard cases.} 

The numerical implementation of the hardening process is switched off whenever the triplet, according to the stability criterion of \citet{Mardling2001}, is no longer hierarchical. When this occurs, in fact, the system dynamics is dominated by chaotic three-body interactions, and the hardening recipe described by eqs. \ref{eq:a_evolution} and \ref{eq:e_evolution}, derived for isolated binary systems, is no longer valid.

We assume that the center of the stellar potential always coincides with the triplet center of mass. Physically, one expects the innermost stellar distribution to adjust, following the dominating MBH gravitational influence. Practically, the code moves the triplet center of mass to the origin of the reference system, where the stellar distribution is centred. The process is replicated every 1000 time-steps (about every one hundred orbits of the inner binary). In the following we discuss the triplet+stellar bulge system, assuming four different values of the initial relative inclination between the inner and the outer 
binary. 
\begin{figure*}
 \begin{minipage}[b]{0.47\textwidth}
   \centering
   \includegraphics[scale=0.4]{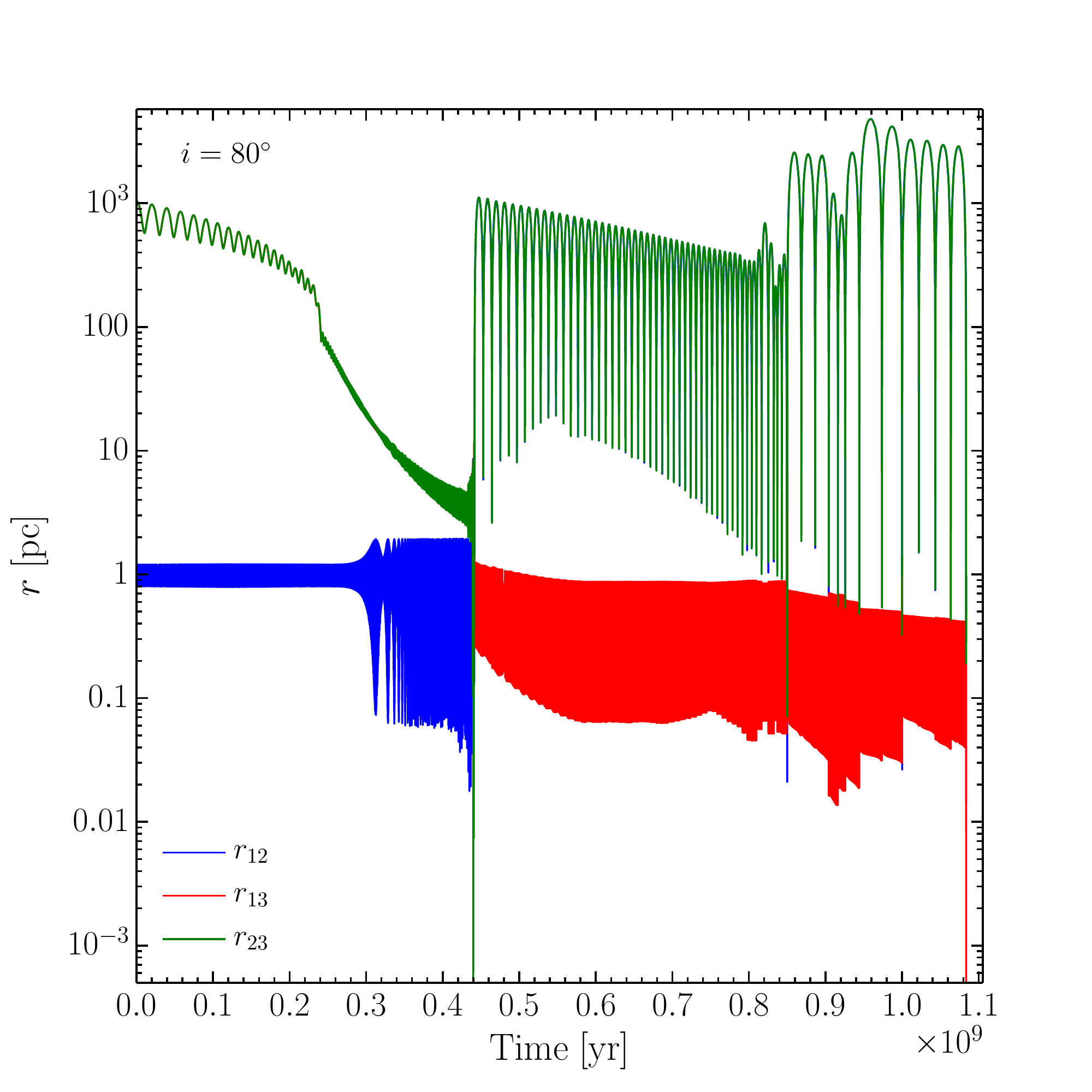}
 \end{minipage}
 \ \hspace{2mm} \
 \begin{minipage}[b]{0.47\textwidth}
  \centering
   \includegraphics[scale=0.4]{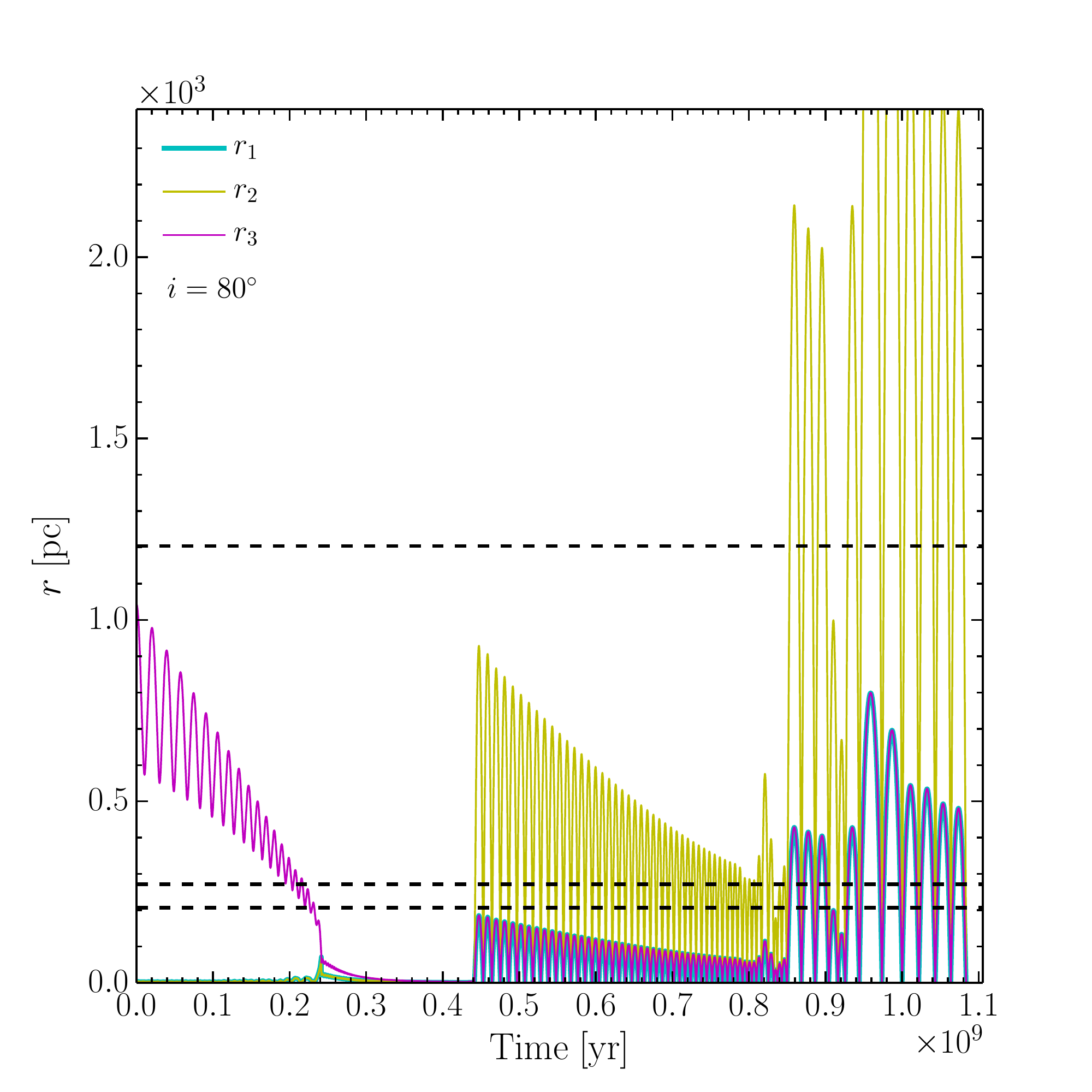}
 \end{minipage}
 \caption{Same as fig.~\ref{fig:test_all_1} but assuming an initial inclination $i=80^\circ$. {\it Left panel}: relative separations. {\it Right panel}: distance of the MBHs from the center of the stellar distribution. The dashed black lines represent, from top to the bottom, the bulge scale radius, the bulge core radius, and the 
 influence radius of the ``original" inner binary (i.e., $m_1+m_2$), respectively. Note in both panels the exchange event between $m_2$ and $m_3$ occurring at $t\simeq 440$ Myr.}  
 \label{fig:test_all_3}
\end{figure*}

In fig.~\ref{fig:test_all_1} we plot, as a function of time from the start of the simulation, 
the relative separation between the three MBH pairs ($r_{12}$ in blue, $r_{13}$ in red, and $r_{23}$ in green) assuming for the 
inclination a value of $i=89^\circ$. 
The evolution can be described by an initial phase lasting $\approx 230$ Myr when the intruder sinks because of dynamical friction, while the 
inner binary is basically unperturbed. Then a bound triplet forms, and the outer binary keeps shrinking because of stellar hardening. At $\approx 300$ Myr the outer binary has shrunk enough to excite K-L oscillations in the inner binary, as clearly shown by the huge periodic variations of $r_{12}$. 
The K-L mechanism is so effective in periodically increasing $e_{\rm in}$, that after $t=439$ Myr from the start of the simulation the inner binary coalesces because of GW emission. It is worth noticing that, from the point of view of $m_3$, the inner binary is essentially a point mass, so that 
$r_{13}\simeq r_{23}$ (i.e., the green and red lines are coincident). Note also that single orbits of the inner binary are not recognisable in the figure. The relative inclination of the two binaries is plotted in fig.~\ref{fig:inc_vs_t89}, upper red curve, as a function of time, clearly showing the oscillations that occur on the (reducing) K-L timescale. 

\begin{figure}
\includegraphics[width=0.47\textwidth]{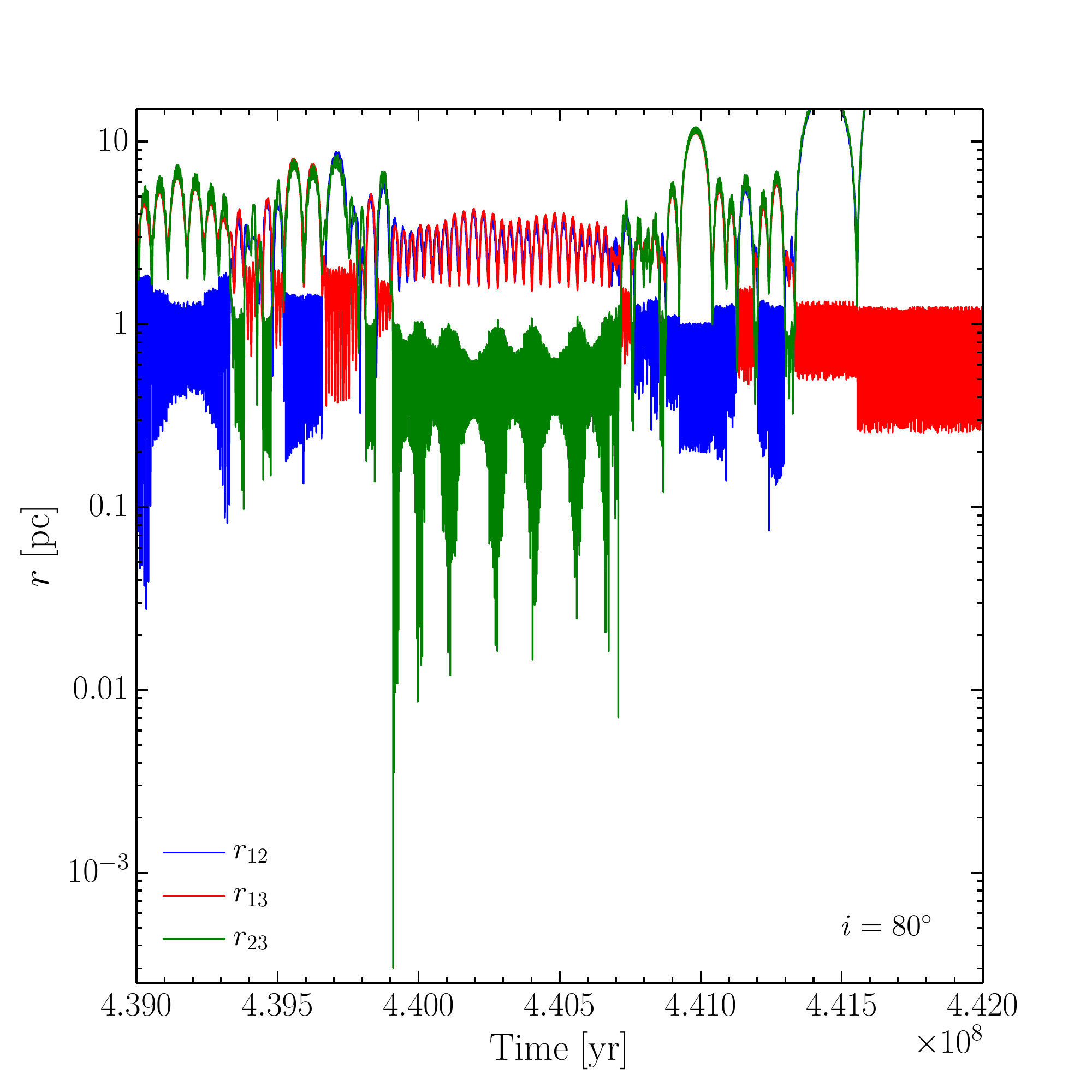}
\caption{Same as left panel of fig.~\ref{fig:test_all_3}, zooming in the exchange event occuring at $t\simeq 440$ Myr.} 
\label{fig:test_all_scambio}
\end{figure}

A further interesting case is shown in fig.~\ref{fig:test_all_3}, where we set $i=80^\circ$. Though the inclination is large enough to excite the K-L oscillations 
(as clearly shown by the blue line in the left panel), the increase of $e_{\rm in}$ is sensitively lower compared to the $i=89^\circ$ case (note the different y-axis scale 
in fig.~\ref{fig:test_all_1}). This allows the outer binary to shrink more since the hardening process can operate for a longer time, undermining 
the secular stability of the triplet. Indeed, at $t\simeq 440$ Myr, an exchange event between $m_2$ and $m_3$ occurs, as seen by the sudden appearance of the red line, i.e., the inner binary is now $m_1+m_3$. The exchange complexity is clearly shown in the time zoom-in of the event (see fig.~\ref{fig:test_all_scambio}). As a consequence of the exchange, $m_2$ is kicked on a very eccentric and much wider orbit, while at the same time the inner binary is relatively stable for the next $\simeq 500$ Myr. At this point a further close encounter with $m_2$ forces the inner binary's eccentricity to greatly increase, then leading to coalescence after $\simeq 1$ Gyr from the start of the simulation. 
We must point out that in this particular case all the MBHs experience 
almost radial oscillations of large amplitude. This can be seen in the right panel of fig.~\ref{fig:test_all_3}, where we plot the time evolution of the 
distance of each of the three MBHs from the center of the stellar distribution.

\begin{figure*}
 \begin{minipage}[b]{0.47\textwidth}
   \centering
   \includegraphics[scale=0.4]{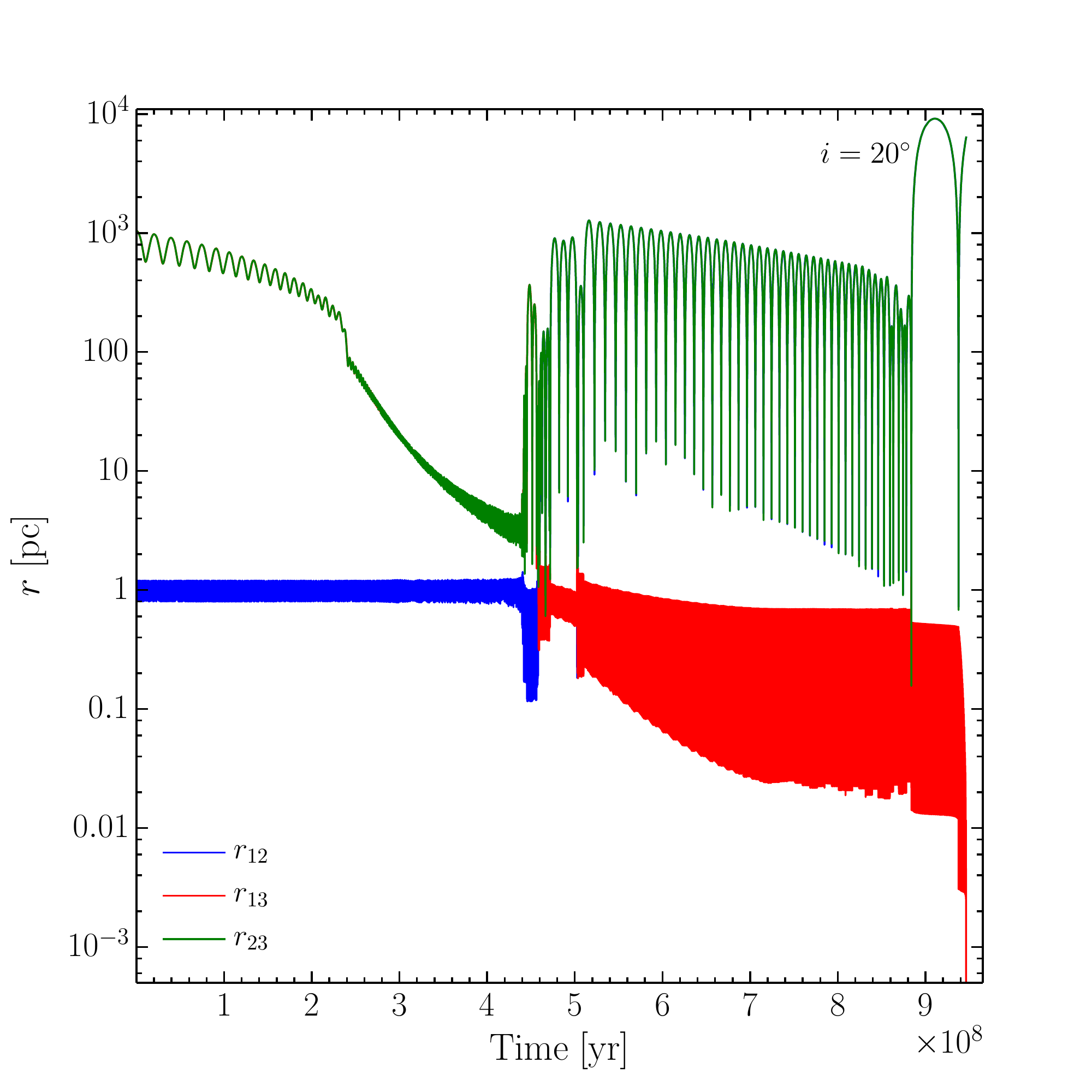}
 \end{minipage}
 \ \hspace{2mm} \
 \begin{minipage}[b]{0.47\textwidth}
  \centering
   \includegraphics[scale=0.4]{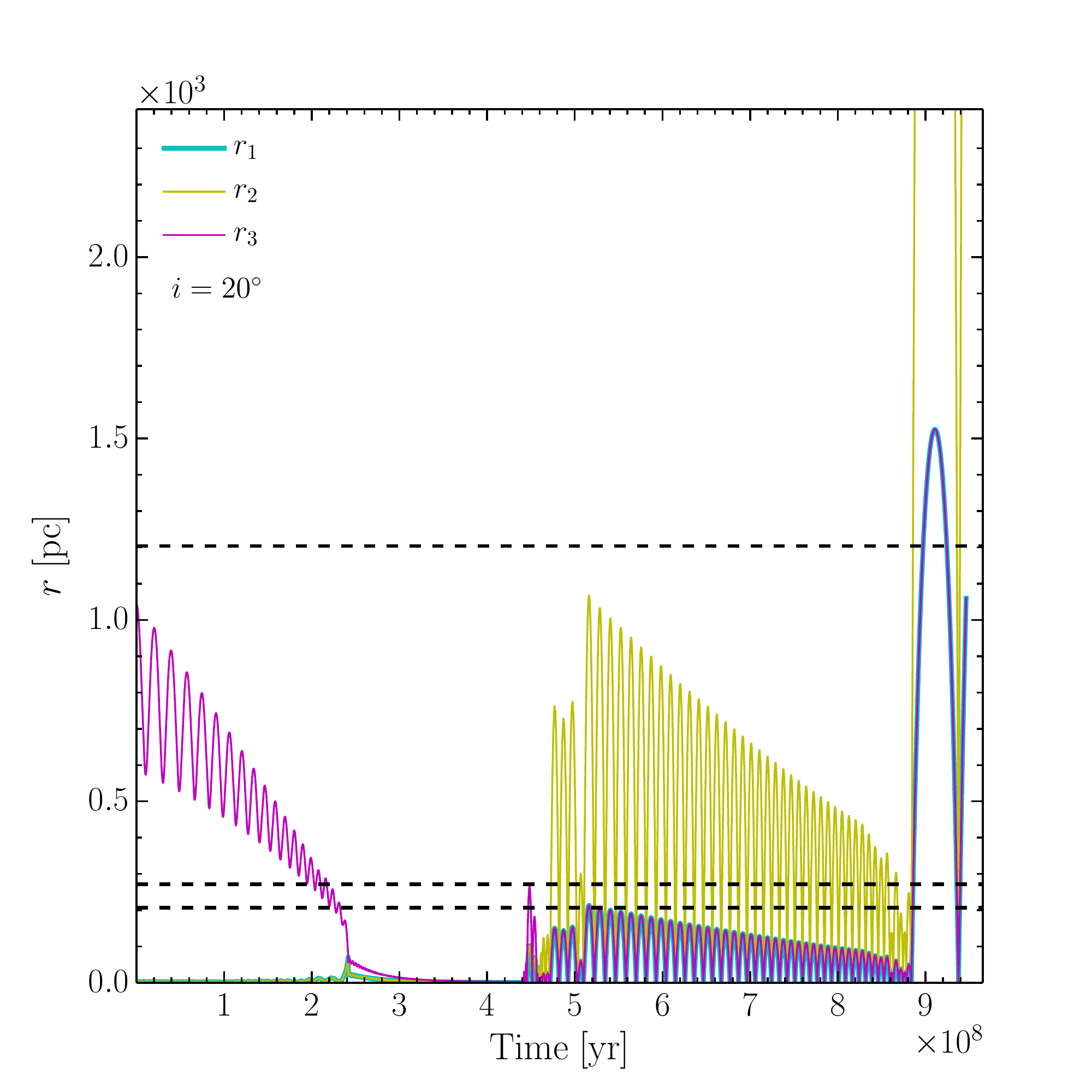}
 \end{minipage}
 \caption{Same as fig.~\ref{fig:test_all_3} but assuming an initial inclination $i=20^\circ$.}  
 \label{fig:test_all_2}
\end{figure*}

\begin{figure}
   \includegraphics[width=0.47\textwidth]{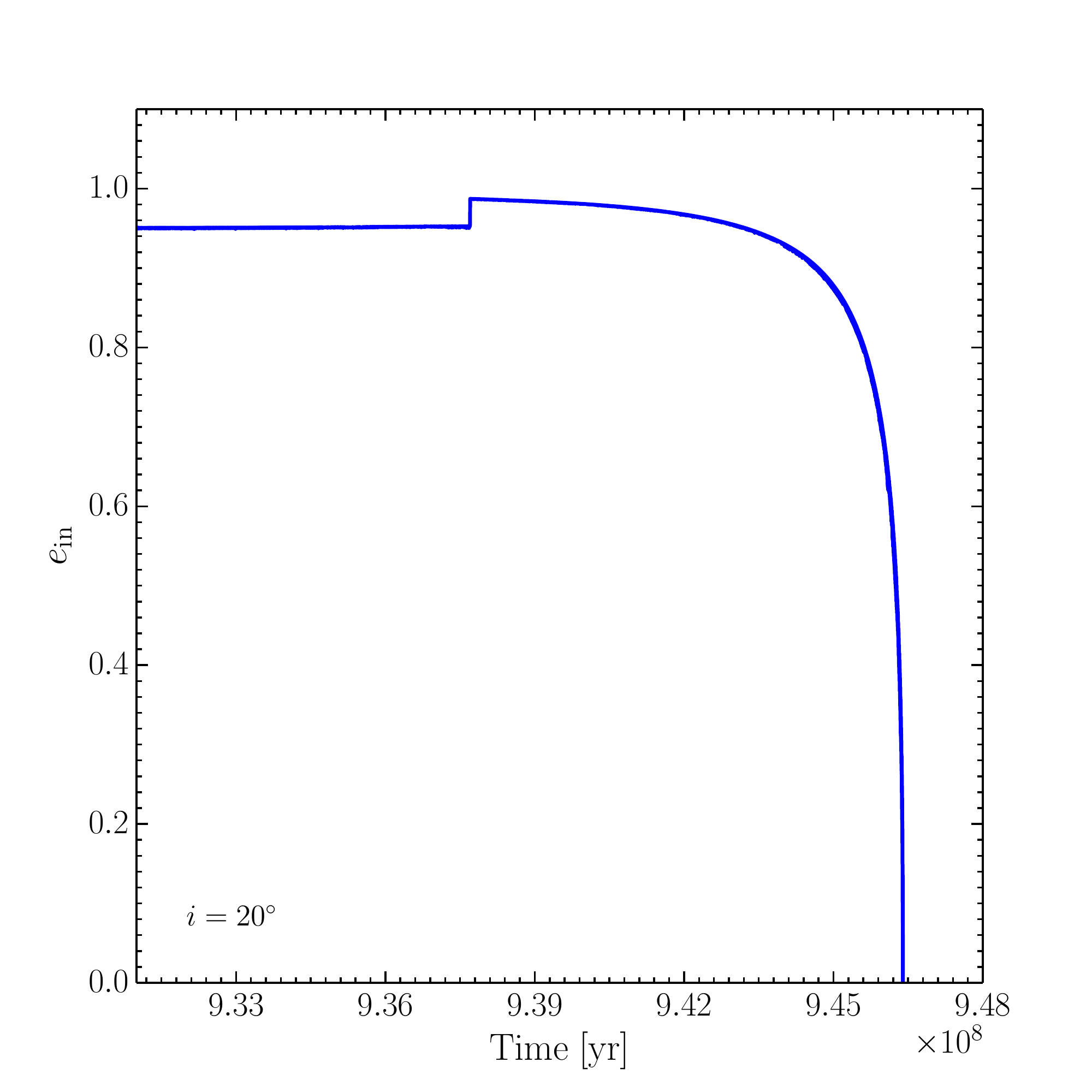}
   \caption{The final phase of the $i=20^\circ$ case. A fly-by of $m_2$ at $t\simeq 937$ Myr causes $e_{\rm in}$ to grow form $\simeq 0.95$ to $\simeq 0.99$, leading the inner binary to coalesce within the next $\simeq 10$ Myr.}
\label{fig:ecc20}
\end{figure}

A case with $i=20^\circ$ is shown in fig.~\ref{fig:test_all_2}. K-L oscillations are not excited since the inclination is below the nominal 
threshold of $\simeq 39^\circ$. The ``original" inner binary is not going to coalesce, then. However, after $\simeq 420$ Myr, $m_3$ and the inner binary experience an energetic close encounter whose final outcome is an exchange between $m_2$ and $m_3$. While $m_2$ is kicked on a 
very eccentric and much wider orbit, the inner binary (now $m_1+m_3$, shown again by the red line) is relatively stable for the next $\simeq 500$ Myr. At this point a fly-by of $m_2$ (which is on a very eccentric, shrinking orbit)  
forces the inner binary's eccentricity to greatly increase, then leading to coalescence within the next $\simeq 10$ Myr (see fig.~\ref{fig:ecc20}).  Also in this case, the three MBHs are slingshot on almost radial orbits in the stellar potential. Indeed, the coalescence of the inner binary occurs when it lies at $\simeq 1$ kpc from the center of the stellar distribution.  

We must note that, whenever the MBHs are forced on almost radial kpc-scale orbits as in in the last two cases discussed, the possible triaxiality of a more realistic stellar distribution could alter the dynamics of the triplet, possibly delaying any close encounter between $m_2$ and the inner binary. We plan to include and analyse the effects of triaxiality in the next implementation of our code. 

We finally analyse a system with relative binary inclination set to $i=10^\circ$. Despite of the low initial inclination, the inner binary is bound 
to coalesce after $t=458$ Myr, as shown in fig.~\ref{fig:test_all_4}. After $\simeq 440$ Myr, the triplet evolution is characterised by many close encounters (we also witness four exchanges) that increase the relative inclination above the K-L critical angle. The final outcome is most probably determined by K-L oscillations with contribution from higher orders \citep[see][]{Li2014}. Note that during the last Myr before the binary merger, the pericenter is as small as $\simeq 1$ mpc, thus making the system a suitable candidate for a PTA burst-like signal. 

The time evolution of the relative inclination of the two binaries is shown in fig.~\ref{fig:inc_vs_t89}, lower blue curve. 
As the intruder gets close enough to the inner binary, small periodic variations of $i$ on long timescales (most probably led by high order K-L resonances) are excited. 
It is only when the three bodies experience the final 
close encounters at $t\simeq 440$ Myrs (eventually leading to the coalescence of the inner binary) that the rapid 
changes in inclination become large and erratic.

\section{Discussion and conclusions}
\subsection{Relevance of the dynamical ingredients included in the code}

\begin{figure*}
 \begin{minipage}[b]{0.47\textwidth}
   \centering
   \includegraphics[scale=0.4]{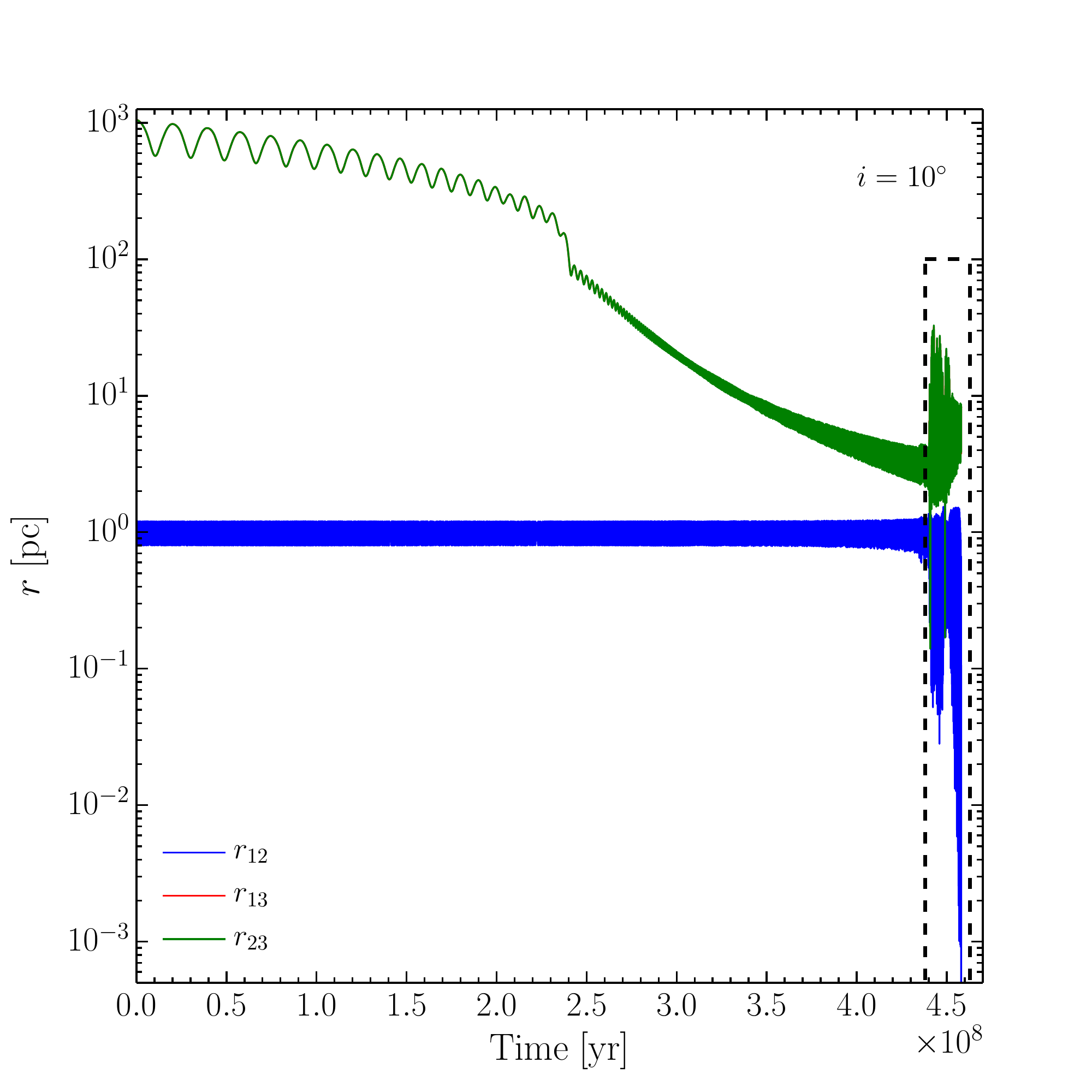}
 \end{minipage}   
 \ \hspace{2mm} \
 \begin{minipage}[b]{0.47\textwidth}
  \centering
   \includegraphics[scale=0.4]{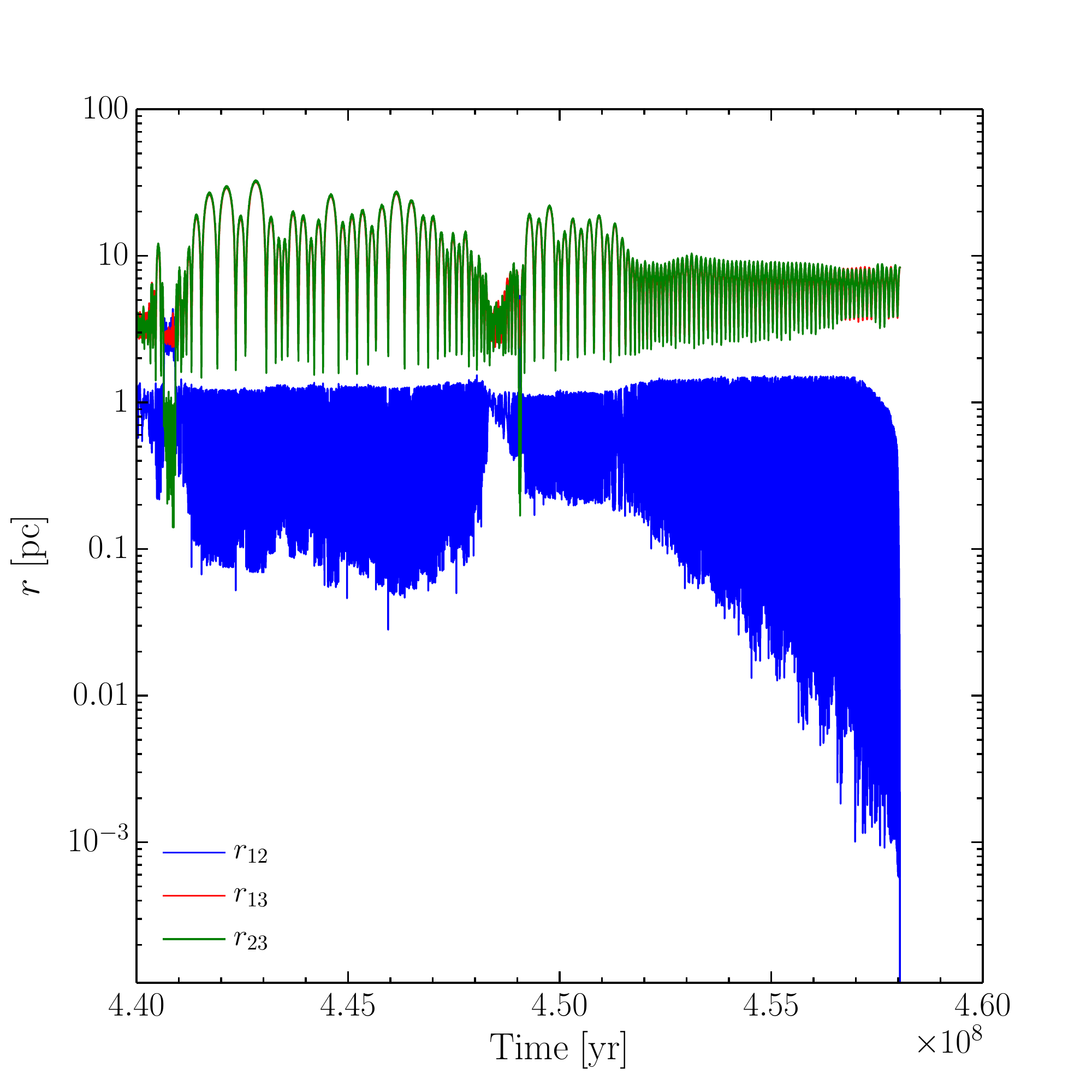}
 \end{minipage}
 \caption{Same as fig.~\ref{fig:test_all_1} but assuming $i=10^\circ$. {\it Left}: entire run. The vertical dashed lines frame the time interval zoomed in the right panel. {\it Right}: the 20 Myr time zoom.}
 \label{fig:test_all_4}
\end{figure*}

The test runs shown in the previous section reveal an extremely complex and diverse phenomenology, in which all the ingredients included in the modelling play a relevant role. Stellar hardening is crucial in bringing the intruder first down to a separation where K-L resonances can be effectively excited ($\sim 10$ pc for the specific masses examined), and then further down to experience strong interactions with individual components of the inner binary. K-L resonances require that the properties of the inner binary do not change over a time $t_{\rm KL}$ (eq.~\ref{eq:TKozai}). Although this is a safe condition for a Newtonian system, it easily breaks down when general relativistic effects are included. In fact, as firstly demonstrated by \cite{Blaes2002}, relativistic induced precession at 1PN and 2PN order can suppress K-L resonances. Considering only the leading 1PN order for simplicity, a comparison between the K-L and relativistic precession rates yields \citep{2011ApJ...729...13C}

\begin{equation}
  \frac{\dot{\omega}_{\rm KL}}{\dot{\omega}_{\rm GR}} \sim \frac{m_3}{m_1+m_2}\left(\frac{a_{\rm in}}{a_{\rm out}}\right)^3\frac{a_{\rm in}}{R_{\rm S,in}}\frac{(1-e_{\rm in}^2)^{1/2}}{(1-e_{\rm out}^2)^{3/2}},
  \label{precession}
\end{equation}

where $R_{\rm S,in}=G(m_1+m_2)/c^2$ is the ``equivalent'' gravitational radius of a MBH with mass equal to the mass of the inner binary. The K-L mechanism is effective only when $\dot{\omega}_{\rm KL}>\dot{\omega}_{\rm GR}$. For example, given a specific inner binary, the intruder has to shrink to a small enough $a_{\rm out}$ to trigger it. This is what we have shown in figures \ref{fig:triplet1} and \ref{fig:triplet2}, where the system with smaller $a_{\rm out}$ (figure \ref{fig:triplet1}) undergoes more effective K-L cycles. Note how the GW driven hardening of the inner binary progressively suppresses the effect. This is because $a_{\rm in}$ decreases, bringing down the $\dot{\omega}_{\rm KL}/\dot{\omega}_{\rm GR}$ ratio in eq.~\ref{precession}. The exact point at which K-L becomes effective also depends on the mass ratio of the intruder with respect to the inner binary, and on the eccentricities of the inner vs. the outer binary. Note that as $e_{\rm in}$ increases, the ratio $\dot{\omega}_{\rm KL}/\dot{\omega}_{\rm GR}$ decreases. Therefore, the K-L mechanism might eventually self-regulate itself: as it grows the inner binary more eccentric, it also makes relativistic precession more effective. The condition $\dot{\omega}_{\rm KL}>\dot{\omega}_{\rm GR}$ might therefore not be satisfied any longer, thus suppressing the K-L effect and eventually altering the overall dynamics of the system. 

Although our investigation is similar in spirit to that of \cite{Hoffman2007}, the above discussion highlights the importance of the differences in the two implementations. In particular, \cite{Hoffman2007} did not include 1PN and 2PN relativistic precession in their equations. This might significantly alter the overall statistical properties of merging binaries (e.g., eccentricities, coalescence timescales) because, as we just discussed, the conditions for triggering K-L resonances, and thus the general dynamics of the system, are different. Moreover, our treatment of the hardening in the stellar background is more accurate, since it also reproduces the eccentricity evolution in the hardening phase. This is important because i), it affects the triggering point of K-L cycles (eq.~\ref{precession}), and ii), it has a strong impact on the stability of the hierarchical triplet and on the probability of close encounters leading to chaotic behaviour. 


\subsection{Astrophysical implications}
The main aim of this paper is to present and validate the code, and a thorough analysis of the dynamics of MBH triplets is deferred to forthcoming papers of the series we planned. Still, the few cases examined here already provide some interesting astrophysical insights that we briefly outline in the following.

In all our simulations, two of the MBHs (not necessarily those originally forming the inner binary) coalesce in less than 1Gyr from the start of the dynamical friction phase of the intruder. Therefore, triple interactions might provide a viable channel to merge MBH binaries in massive, low density galaxies, where hardening against the stellar and gaseous background might act on a timescale of several Gyr \citep{2011ApJ...732...89K,2015ApJ...810...49V,Sesana2015}. At least a fraction of massive elliptical galaxies,
which host the most massive binaries targeted by PTAs, might realistically undergo multiple mergers at $z\lsim 1$ \citep[see][and references therein]{2015MNRAS.446...38G}. If the typical coalescence timescale of the formed MBHBs is several Gyr, then the occurrence of a second merger will bring in an intruder, typically leading to coalescence on a much shorter timescale. Triple interactions might therefore be an important channel for merging very massive, gas-poor low redshift binaries. Whether it is also important for lower mass system at higher redshifts is less clear. The larger availability of cold gas, together with extremely high density environments \citep{2016arXiv160400015K} might in fact result in more efficient coalescences on a timescale $\lsim 10^8$yr. Note, however, that mergers are way more common at high $z$, and that the relevant dynamical timescales are also much shorter. If seed black holes are abundant in high redshift protogalaxies, it is then possible that many triple systems will form following subsequent galaxy mergers.

The dynamics of individual triplets presents in itself extremely interesting features. For example, the system shown in fig.~\ref{fig:test_all_3} undergoes a close encounter at $T\approx 440$Myr, with a closest passage at $\lsim 10^{-3}$ pc. On the other hand, all the other cases show a late phase when $e_{\rm in}\gsim 0.99$, eventually promoting the final coalescence of the inner binary. In particular, in fig.~\ref{fig:test_all_4} we see that the inner system spends its last million years of life with $e_{\rm in}\approx0.999$ before coalescence. If this behaviour is general, then it carries important consequences for the GW signals expected from these sources. Indeed, \cite{Pau2010} showed that a very eccentric binary emits relatively broad-band bursts centred at frequency $f\propto[a(1-e)]^{-3/2}$ (which is the frequency of a circular binary with semimajor axis equal to the periastron of the eccentric system), which we can parametrise as \citep[see][]{Wen2003,Antonini2016}
\begin{equation}
  f \approx 4 \times10^{-7}{\rm Hz}\left(\frac{M}{10^8\msun}\right)^{-1}\left(\frac{\alpha}{100}\right)^{-3/2}.
  \label{frequency}
\end{equation}

Here $\alpha=r_p/R_{\rm S,in}$, where $r_p$ is the binary periastron. Note that, for $M=10^8 M_\odot$, $\alpha=100$ corresponds to $r_p \approx 1$ mpc, which is the typical value found in our test cases. Therefore, right before coalescence or during extremely close encounters (such as the one shown in fig.~\ref{fig:test_all_3}), these systems will emit intense bursts of gravitational radiation of the duration of $\approx 1$ month (for the masses considered here), which might be detectable by PTAs. Moreover, if binaries typically coalesce with resonance-induced high eccentricities (as in most of the cases shown here), there might be other profound consequences for the overall GW signal that PTAs are hunting. In the most extreme scenario, very high eccentricities will dramatically suppress the low frequency signal, shifting most of the emitted power at higher frequencies. Moreover, the statistical properties of the signal might look quite different, featuring a collection of burst-like events of duration of months-to-years, rather than the superposition of continuous periodic sources. Although eq.~\ref{frequency} gives the central frequency of the signal, the burst is expected to be broad-band, possibly extending to frequencies  more than an order of magnitude higher (for $e\gsim 0.99$). Although \cite{Pau2010} found that massive ($M\gsim 10^8 \msun$) systems are extremely unlikely to burst in the {\it eLISA} band, if triplets are also common among low mass systems in the high redshift Universe, burst-like signals may also be a relatively frequent occurrence in the {\it eLISA} band.

\subsection{Outlook}
In oder to quantify the importance of the astrophysical consequences sketched in the previous subsection, an extensive parameter space study is in progress, covering the relevant mass, mass ratio and eccentricity range of astrophysical triplets in a cosmological frame of structure formation and evolution. 

With this goal in mind, we devoted this first paper to the description and validation of the main code. We included all PN terms in the three-body equations of motion up to 2.5PN order. We also included a simple prescription for dynamical friction, and an ad-hoc designed fictitious force that reproduces both the semi-major axis and the eccentricity evolution of MBHB hardening in a stellar background. The effect of the stellar background itself is included in the equations of motion, providing additional Newtonian precession. We tested the stability of the code with a number of standard tests and by comparing the dynamical outcome of rather complex situations to results obtained by other groups with the ARCHAIN code, finding good agreement. We also tested the importance of non-dissipative 1PN and 2PN terms in the dynamical evolution of the system, showing how they alter the excitation of K-L resonances.

Our code includes most of the physics relevant to the dynamics of massive triplets in stellar systems, can be easily expanded to include further dynamical features, such as the effects of a non-spherical potential, and is versatile enough to allow an efficient exploration of the parameter space relevant to astrophysical triplets.  This will provide the necessary statistics of close encounter to assess the occurrence of individual GW bursts, and will allow us to properly construct the eccentricity distribution of the systems approaching coalescence, to assess the implications for GW detection with {\it eLISA} and PTAs.

\section*{Acknowledgements}
We have benefitted from many invaluable discussions with Monica Colpi and Massimo Dotti on various aspects of BH dynamics.
We also thank Guillaume Faye for insightful clarifications on the three-body PN Hamiltonian, and the anonymous referee for comments and suggestions that helped us to improve the paper. MB and FH acknowledge partial financial support from the INFN TEONGRAV specific initiative. EB acknowledges support from the European Union's Seventh Framework Programme (FP7/PEOPLE-2011-CIG) through the Marie Curie Career Integration Grant GALFORMBHS PCIG11-GA-2012-321608, and from the H2020-MSCA-RISE-2015 Grant No. StronGrHEP-690904. AS is supported by a University Research Fellowship of the Royal Society. This work has made use of the Horizon Cluster, hosted by the Institut d'Astrophysique de Paris. We thank Stephane Rouberol for running smoothly this cluster for us.

\bibliographystyle{mn2e}
\bibliography{biblio} 

\onecolumn
\appendix
\section{Hamiltonians}

We report here the three-body Newtonian, 1PN and 2PN Hamiltonians of \citet{Galaviz2011}, where we corrected a typo in the third to last 
line of the $H_2$ term reported below ($r_{\alpha\beta}^2 \rightarrow r_{\alpha\beta}$). 

\begin{equation}
H_0 = \dfrac{1}{2}\sum_{\alpha}\dfrac{|\vec{p}_{\alpha}|^2}{m_{\alpha}} - \dfrac{G}{2}\sum_{\alpha}\sum_{\beta \neq \alpha} \dfrac{m_{\alpha}m_{\beta}}{r_{\alpha\beta}}
\end{equation}

\begin{equation}
\begin{split}
H_1 &= -\dfrac{1}{8}\sum_{\alpha} m_{\alpha}\left(\dfrac{|\vec{p}_{a}|^2}{m_{\alpha}^2}\right)^2
-\dfrac{G}{4}\sum_{\alpha}\sum_{\beta \neq \alpha}\dfrac{1}{r_{\alpha\beta}}\biggl[ 6\dfrac{m_{\beta}}{m_{\alpha}}|\vec{p}_{\alpha}|^2
-7\vec{p}_{\alpha}\cdot\vec{p}_{\beta} - (\vec{n}_{\alpha\beta}\cdot\vec{p}_{\alpha})(\vec{n}_{\alpha\beta}\cdot\vec{p}_{\beta}) \biggr]
+\dfrac{G^2}{2}\sum_{\alpha}\sum_{\beta \neq \alpha}\sum_{\gamma \neq \alpha} \dfrac{m_{\alpha}m_{\beta}m_{\gamma}}{r_{\alpha\beta}r_{\alpha\gamma}}
\end{split}
\end{equation}

\begin{equation}
\begin{split}
H_2 &= \dfrac{1}{16}\sum_{\alpha} m_{\alpha} \left(\dfrac{|\vec{p}_{\alpha}|^2}{m_{\alpha}^2}\right)^3 
+\dfrac{G}{16}\sum_{\alpha}\sum_{\beta \neq \alpha} \dfrac{(m_{\alpha} m_{\beta})^{-1}}{r_{\alpha\beta}} \biggl[ 10 \left(\dfrac{m_{\beta}}{m_{\alpha}}|\vec{p}_{\alpha}|^2\right)^2 - 11 |\vec{p}_{\alpha}|^2 |\vec{p}_{\beta}|^2 - 2(\vec{p}_{\alpha}\cdot\vec{p_{\beta}})^2 \\
&+10|\vec{p}_{\alpha}|^2 (\vec{n}_{\alpha\beta}\cdot\vec{p}_{\beta})^2 - 12(\vec{p}_{\alpha}\cdot\vec{p}_{\beta})(\vec{n}_{\alpha\beta}\cdot\vec{p}_{\alpha})(\vec{n}_{\alpha\beta}\cdot\vec{p}_{\beta}) - 3(\vec{n}_{\alpha\beta}\cdot\vec{p}_{\alpha})^2(\vec{n}_{\alpha\beta}\cdot\vec{p}_{\beta})^2 \biggr] \\
&+\dfrac{G^2}{8}\sum_{\alpha}\sum_{\beta \neq \alpha }\sum_{\gamma \neq \alpha} \dfrac{1}{r_{\alpha\beta}r_{\alpha\gamma}} \biggl[ 18\dfrac{m_{\beta} m_{\gamma}}{m_{\alpha}}|\vec{p}_{\alpha}|^2 + 14\dfrac{m_{\alpha} m_{\gamma}}{m_{\beta}}|\vec{p}_{\beta}|^2 - 2\dfrac{m_{\alpha} m_{\gamma}}{m_{\beta}}(\vec{n}_{\alpha\beta}\cdot \vec{p}_{\beta})^2\\
&- 50 m_{\gamma} (\vec{p}_{\alpha}\cdot\vec{p}_{\beta}) + 17 m_{\alpha} (\vec{p}_{\beta}\cdot\vec{p}_{\gamma}) - 14 m_{\gamma} (\vec{n}_{\alpha\beta}\cdot\vec{p}_{\alpha})(\vec{n}_{\alpha\beta}\cdot\vec{p}_{\beta})\\
&+ 14 m_{\alpha} (\vec{n}_{\alpha\beta}\cdot\vec{p}_{\beta})(\vec{n}_{\alpha\beta}\cdot\vec{p}_{\gamma}) + m_{\alpha} (\vec{n}_{\alpha\beta}\cdot\vec{n}_{\alpha\gamma})(\vec{n}_{\alpha\beta}\cdot\vec{p}_{\beta})(\vec{n}_{\alpha\gamma}\cdot\vec{p}_{\gamma}) \biggr]\\
&+\dfrac{G^2}{8}\sum_{\alpha} \sum_{\beta\neq \alpha}\sum_{\gamma\neq \alpha} \dfrac{1}{r_{\alpha\beta}^2} \biggl[ 2m_{\beta}(\vec{n}_{\alpha\beta}\cdot\vec{p}_{\alpha})(\vec{n}_{\alpha\gamma}\cdot\vec{p}_{\gamma}) + 2m_{\beta}(\vec{n}_{\alpha\beta}\cdot\vec{p}_{\beta})(\vec{n}_{\alpha\gamma}\cdot\vec{p}_{\gamma})\\
&+ \dfrac{m_{\alpha} m_{\beta}}{m_{\gamma}} \bigl( 5(\vec{n}_{\alpha\beta}\cdot\vec{n}_{\alpha\gamma})|\vec{p}_{\gamma}|^2 - (\vec{n}_{\alpha\beta}\cdot\vec{n}_{\alpha\gamma})(\vec{n}_{\alpha\gamma}\cdot\vec{p}_{\gamma})^2 - 14(\vec{n}_{\alpha\beta}\cdot\vec{p}_{\gamma})(\vec{n}_{\alpha\gamma}\cdot\vec{p}_{\gamma}) \bigr) \biggr]\\
&+\dfrac{G^2}{4}\sum_{\alpha} \sum_{\beta \neq \alpha} \dfrac{m_{\alpha}}{r_{\alpha\beta}^2} \biggl[ \dfrac{m_{\beta}}{m_{\alpha}}|\vec{p}_{\alpha}|^2 + \dfrac{m_{\alpha}}{m_{\beta}}|\vec{p}_{\beta}|^2 - 2(\vec{p}_{\alpha}\cdot\vec{p}_{\beta}) \biggr]\\
&+\dfrac{G^2}{2}\sum_{\alpha}\sum_{\beta\neq \alpha}\sum_{\gamma\neq \alpha,\beta} \dfrac{(n_{\alpha\beta}^i+n_{\alpha\gamma}^i)(n_{\alpha\beta}^j+n_{\gamma\beta}^j)}{(r_{\alpha\beta}+r_{\beta\gamma}+r_{\gamma\alpha})^2} \biggl[ 8m_{\beta}(p_{\alpha i}p_{\gamma j}) - 16m_{\beta}(p_{\alpha j}p_{\gamma i})\\
&+ 3m_{\gamma}(p_{\alpha i}p_{\beta j}) + 4 \dfrac{m_{\alpha} m_{\beta}}{m_{\gamma}} (p_{\gamma i}p_{\gamma j}) + \dfrac{m_{\beta} m_{\gamma}}{m_{\alpha}}(p_{\alpha i}p_{\alpha j})\biggr]\\
&+\dfrac{G^2}{2}\sum_{\alpha}\sum_{\beta\neq \alpha}\sum_{\gamma\neq \alpha,\beta} \dfrac{m_{\alpha}m_{\beta}m_{\gamma}}{(r_{\alpha\beta}+r_{\beta\gamma}+r_{\gamma\alpha})r_{\alpha\beta}} \biggl[ 8\dfrac{\vec{p}_{\alpha}\cdot\vec{p}_{\gamma} - (\vec{n}_{\alpha\beta}\cdot\vec{p}_{\alpha})(\vec{n}_{\alpha\beta}\cdot\vec{p}_{\gamma})}{m_{\alpha} m_{\gamma}}\\
&- 3\dfrac{ \vec{p}_{\alpha}\cdot\vec{p}_{\beta} - (\vec{n}_{\alpha\beta}\cdot\vec{p}_{\alpha})(\vec{n}_{\alpha\beta}\cdot\vec{p}_{\beta}) }{m_{\alpha} m_{\beta}} - 4\dfrac{|\vec{p}_{\gamma}|^2 - (\vec{n}_{\alpha\beta}\cdot\vec{p}_{\gamma})^2 }{m_{\gamma}^2} - \dfrac{|\vec{p}_{\alpha}|^2 - (\vec{n}_{\alpha\beta}\cdot\vec{p}_{\alpha})^2 }{m_{\alpha}^2} \biggr]\\
&-\dfrac{G^3}{2}\sum_{\alpha}\sum_{\beta \neq \alpha}\biggl(\sum_{\gamma \neq \alpha,\beta} \dfrac{m_{\alpha}^2 m_{\beta} m_{\gamma}}{r_{\alpha\beta}^2 r_{\beta\gamma}} +\dfrac{1}{2}\sum_{\gamma \neq \beta} \dfrac{m_{\alpha}^2 m_{\beta} m_{\gamma}}{r_{\alpha\beta}^2 r_{\beta\gamma}} \biggr)\\
&-\dfrac{3G^3}{8}\sum_{\alpha}\sum_{\beta \neq \alpha}\biggl(\sum_{\gamma \neq \alpha} \dfrac{m_{\alpha}^2 m_{\beta} m_{\gamma}}{r_{\alpha\beta}^2 r_{\alpha\gamma}} + \sum_{\gamma \neq \alpha,\beta} \dfrac{m_{\alpha}^2 m_{\beta} m_{\gamma}}{r_{\alpha\beta}^2 r_{\beta\gamma}} \biggr)\\
&-\dfrac{3G^3}{8}\sum_{\alpha}\sum_{\beta \neq \alpha}\sum_{\gamma \neq \alpha,\beta} \dfrac{m_{\alpha}^2 m_{\beta} m_{\gamma}}{r_{\alpha\beta} r_{\alpha\gamma} r_{\beta\gamma}}\\
&-\dfrac{G^3}{64}\sum_{\alpha}\sum_{\beta \neq \alpha}\sum_{\gamma \neq \alpha,\beta} \dfrac{ m_{\alpha}^2 m_{\beta} m_{\gamma} }{ r_{\alpha\beta} r_{\alpha\gamma}^3 r_{\beta\gamma} } \biggl[ 18 r_{\alpha\gamma}^2 - 60 r_{\beta\gamma}^2 -24 r_{\alpha\gamma}(r_{\alpha\beta}+r_{\beta\gamma})
+ 60 \dfrac{r_{\alpha\gamma} r_{\beta\gamma}^2}{r_{\alpha\beta}} + 56 r_{\alpha\beta}r_{\beta\gamma} - 72 \dfrac{r_{\beta\gamma}^3}{r_{\alpha\beta}} + 35\dfrac{r_{\beta\gamma}^4}{r_{\alpha\beta}^2} + 6 r_{\alpha\beta}^2 \biggr]\\
&-\dfrac{G^3}{4}\sum_{\alpha}\sum_{\beta \neq \alpha}\dfrac{m_{\alpha}^2 m_{\beta}^2}{r_{\alpha\beta}^3}.
\end{split}
\end{equation}

\end{document}